\documentclass[11pt,titlepage,twoside,a4paper]{article} 
\date{Version of 1 March 2001}
\title{The Inductive Approach to\\ Verifying Cryptographic Protocols}
\author{Lawrence C. Paulson\\
        Computer Laboratory, University of Cambridge\\[2ex]
  \texttt{lp15@cam.ac.uk}}                               

\usepackage{url,amsmath,epsfig,proof}
\usepackage{ttbox}
\usepackage{basic}
\def\lbb{\mathopen{\{\kern-.30em|}}
\def\rbb{\mathclose{|\kern-.32em\}}}
\def\comp#1{\lbb#1\rbb}

\newcommand\INS[1]{\{#1\}\cup}

\newcommand\Server{\textsf{S}}
\newcommand\Friend{\mathop{\textsf{Friend}}}
\newcommand\Spy{\textsf{Spy}}

\newcommand\Agent{\mathop{\textsf{Agent}}}
\newcommand\Nonce{\mathop{\textsf{Nonce}}}
\newcommand\Number{\mathop{\textsf{Number}}}
\newcommand\Na{\mathit{Na}}
\newcommand\Nb{\mathit{Nb}}
\newcommand\Nc{\mathit{Nc}}
\newcommand\Key{\mathop{\textsf{Key}}}
\newcommand\Ka{\mathit{Ka}}
\newcommand\Kb{\mathit{Kb}}
\newcommand\Kc{\mathit{Kc}}
\newcommand\Kab{\mathit{Kab}}
\newcommand\Kbc{\mathit{Kbc}}
\newcommand\Kcs{\mathit{Kcs}}
\newcommand\Kcd{\mathit{Kcd}}
\newcommand\Kca{\mathit{Kca}}
\newcommand\Hash{\mathop{\textsf{Hash}}\nolimits}
\newcommand\Crypt{\mathop{\textsf{Crypt}}}

\newcommand\parts{\mathop{\textsf{parts}}}
\newcommand\analz{\mathop{\textsf{analz}}}
\newcommand\synth{\mathop{\textsf{synth}}}
\newcommand\keysFor{\mathop{\textsf{keysFor}}}

\newcommand\bad{\mathop{\textsf{bad}}}

\newcommand\SAYS{\mathop{\textsf{Says}}}

\newcommand\Says[3]{\textsf{Says}\,#1\,#2\,#3}

\newcommand\Notes[2]{\textsf{Notes}\,#1\,#2}
\newcommand\spies[1]{\textsf{spies}\,#1}
\newcommand\seespy{\spies{evs}}

\newcommand\initState{\mathop{\textsf{initState}}}
\newcommand\cons{\mathbin{\#}}

\newcommand\used{\mathop{\textsf{used}}}

\newcommand\shrK{\mathop{\textsf{shrK}}}
\newcommand\pubK{\mathop{\textsf{pubK}}}
\newcommand\priK{\mathop{\textsf{priK}}}
\newcommand\otway{\mathop{\textsf{otway}}}
\newcommand\sol{\mathop{\textsf{set}}}

\newcommand\N{\mathbf{N}}
\newcommand\Suc{\mathop{\textsf{Suc}}}
\newcommand\nat{\textsf{nat}}
\newcommand\agent{\textsf{agent}}
\newcommand\msg{\textsf{msg}}

\newcommand\eqdef{\stackrel{\mathrm{def}}{=}}
\newcommand\recur{\mathop{\mathsf{recur}}}
\newcommand\respond{\mathop{\mathsf{respond}}}
\newcommand\responses{\mathop{\mathsf{responses}}}
\let\imp=\longrightarrow

\newbox\RuleLabelBox
\setbox\RuleLabelBox=\hbox{\footnotesize\tt{Fake}}
\newcommand\RuleLabel[1]{\makebox[\wd\RuleLabelBox][l]{\textrm{#1}}}

\begin{document}

\maketitle
\begin{abstract}
  Informal arguments that cryptographic protocols are secure can be made
  rigorous using \emph{inductive definitions}.  The approach is based on
  ordinary predicate calculus and copes with infinite-state systems.  Proofs
  are generated using Isa\-belle/HOL\@.  The human effort required to analyze
  a protocol can be as little as a week or two, yielding a proof script that
  takes a few minutes to run.
  
  Protocols are inductively defined as sets of traces.  A trace is a list of
  communication events, perhaps comprising many interleaved protocol runs.
  Protocol descriptions incorporate attacks and accidental losses.  The model
  spy knows some private keys and can forge messages using components
  decrypted from previous traffic.  Three protocols are analyzed below:
  Otway-Rees (which uses shared-key encryption), Needham-Schroeder (which uses
  public-key encryption), and a recursive protocol~\cite{bull-otway} (which is
  of variable length).
  
  One can prove that event~$ev$ always precedes event~$ev'$ or that
  property~$P$ holds provided $X$ remains secret.  Properties can be proved
  from the viewpoint of the various principals: say, if $A$ receives
  a final message from~$B$ then the session key it conveys is good.
\end{abstract}

\pagenumbering{roman}\tableofcontents\cleardoublepage
\pagenumbering{arabic}
\setcounter{page}{1}

\section{Introduction}

Cryptographic protocols are intended to let agents communicate securely over
an insecure network.  An obvious security goal is \emph{secrecy}: a spy cannot
read the contents of messages intended for others.  Also important is
\emph{authenticity}: if a message appears to be from Alice, then Alice 
sent precisely that message, and any nonces or timestamps within it give a
correct indication of its freshness.  This paper will not discuss denial
of service.

A typical protocol allows~$A$ to make contact with~$B$, delivering a key to
both parties for their exclusive use.  They may involve as few as two
messages, but are surprisingly hard to get right.  One problem is the
combinatorial complexity of the messages that an intruder could generate.  A
quite different problem is to specify precisely what properties the protocol
is intended to achieve.  Anderson and Needham's excellent
tutorial~\cite{anderson-satan} presents several examples and defines the
terminology used below.

Formal methods can be used to analyze security protocols.  Two popular
approaches are {\em state exploration} and {\em belief logics}.
\begin{itemize}
\item State exploration methods~\cite{ryan96} model the protocol as a finite
  state system.  An exhaustive search checks that all reachable states are
  safe.  Lowe uses a general-purpose model-checker,
  FDR~\cite{lowe-fdr,lowe-splice}; the Interrogator~\cite{kmm-three} is a
  specialized tool.  Attacks are quickly found, but keeping the
  state space small requires drastic simplifying assumptions.
  
\item Belief logics formalize what an agent may infer from messages received.
  The original BAN logic~\cite{ban89} allows short, abstract proofs.  It has
  identified some protocol flaws but missed others.  New belief
  logics~\cite{mao-formal} address some weaknesses of BAN but sacrifice its
  simplicity.
\end{itemize}

We can fruitfully borrow from both approaches: from the first, a concrete
notion of events, such as~$A$ sending~$X$ to~$B$; from the second, the idea of
deriving guarantees from each message.  Protocols are formalized as the set of
all possible traces, which are lists of events such as `$A$ sends~$X$ to~$B$.'
An agent may extend a trace in any way permitted by the protocol, given what
he can see in the current trace.  Agents do not know the true sender of a
message and may forward items that they cannot read.  One agent is an active
attacker.

Properties are proved by induction on traces, using the theorem prover
Isabelle~\cite{paulson-isa-book}.  Analyzing a new protocol requires several
days' effort, while exploring the effects of a change to an existing protocol
often takes just a few hours.  Laws and proof techniques developed for one
protocol are often applicable generally.

The approach is oriented around proving guarantees, but their absence can
indicate possible attacks.  In this way, I have discovered an attack on the
variant of the Otway-Rees protocol suggested by Burrows et al.\ts\cite[page
247]{ban89}.  (At the time, I was unaware of Mao and Boyd's earlier
attack~\cite{mao-formal}.)  Even if no attacks are found, the structure of the
proof yields insights into the protocol.

The paper goes on to describe the method, first in overview
(\S\ref{sec:overview}) and then in some detail~(\S\ref{sec:theory}).  Three
protocols are then analyzed.  Otway-Rees illustrates the shared-key
model~(\S\ref{sec:otway}); Needham-Schroeder illustrates the public-key
model~(\S\ref{sec:ns}); the recursive authentication
protocol~\cite{bull-otway} demonstrates how to deal with
$n$-way authentication~(\S\ref{sec:recur}).  Related work is
discussed~(\S\ref{sec:related}) and conclusions
given~(\S\ref{sec:conclusions}).

\section{Overview of the Inductive Method}\label{sec:overview}

Informal arguments for a protocol's correctness are conducted in terms of what
could or could not happen.  Here is a hypothetical dialogue:
\begin{quote}
\emph{Salesman.}  At the end of a run, only Alice and~Bob can possibly know the
session key~$\Kab$.

\emph{Customer.}  What about an eavesdropper?

\emph{Salesman.}  He can't read the certificates without Alice or~Bob's
long-term keys, which he can't get.

\emph{Customer.}  Could an attacker trick Bob into accepting a key shared with
himself?

\emph{Salesman.}  The use of identifying nonces prevents that. 
\end{quote}
The customer may find such arguments unconvincing, but they can be made
rigorous.  The necessary formal tool is the \emph{inductive
  definition}~\cite{aczel77}.  Each inductive definition lists the possible
actions that an agent or system can perform.  The corresponding induction rule
lets us reason about the consequences of an arbitrary finite sequence of such
actions.  Induction has long been used to specify the semantics of programming
languages~\cite{hennessy90}; it copes well with nondeterminism.  (Plotkin
conceived this application of inductive definitions, while Huet pioneered
their use in proof tools.)

For security protocols, the model must specify the capabilities of an attacker.
Several inductively-defined operators are useful.  One
($\parts$) merely returns all the components of a set of messages.  Another
($\analz$) models the decryption of past traffic using available keys.
Another ($\synth$) models the forging of messages.  The attacker is
specified---independently of the protocol!---in terms of $\analz$ and
$\synth$.  Algebraic laws governing $\parts$, $\analz$ and $\synth$ have been
proved by induction and are invaluable for reasoning about protocols.

The inductive protocol definition models the behaviour of honest agents
faithfully executing protocol steps in the presence of the attacker.  It can
even model carelessness, such as agents accidentally revealing secrets.  The
inherent nondeterminism models the possibility of an agent's being
unavailable.

Belief logics allow short proofs; the main reason for mechanizing
them~\cite{brackin-gny} is to eliminate human error.  In contrast, inductive
verification of protocols involves long and detailed proofs.  Each
safety property is proved by induction over the protocol.  Each case considers
a state of the system that might be reached by the corresponding protocol
step.  Simplifying the safety property for that case may reveal a combination
of circumstances leading to its violation.  Only if all cases are covered has
the property been proved.
\begin{quote}
\emph{Customer.}  What's to stop somebody's tampering with the nonce in step~2
and later sending Alice the wrong certificate?

\emph{Salesman.}  Is there somebody less experienced I could talk to?
\end{quote}

\subsection{Messages}\label{sec:messages}

Traditional protocol notation is not ideal for mechanization.  Expressing
concatenation by a comma, as in $A,B$, can be ambiguous; enclosing it in
braces, as in $\{A,B\}$, invites confusion with a two-element set.  The
machine syntax uses fat braces to express concatenation: $\comp{A,B}$.
Informal protocol descriptions omit outer-level braces and indicate encryption
by a notation such as $\comp{\Na,\Kab}_{\Ka}$.

Individual protocol descriptions rest on a common theory of message analysis.
Message items may include
\begin{itemize}
\item agent names~$A$, $B$, \ldots;
\item nonces~$\Na$, $\Nb$, \ldots; 
\item keys~$\Ka$, $\Kb$, $\Kab$, \ldots; 
\item compound messages $\comp{X,X'}$, 
\item hashed messages $\Hash X$,
\item encrypted messages $\Crypt K X$.
\end{itemize}
With public-key encryption, $K^{-1}$ is the inverse of key~$K$.  The equality
$K^{-1}=K$ expresses that $K$ is a symmetric key.  The theory assumes
$(K^{-1})^{-1}=K$ for all~$K$.

Nonces are of two kinds: those that are guessable and those that are not.
Sequence numbers and timestamps can be regarded as guessable, but not 40-byte
random strings.

An encrypted message can neither be altered nor read without the appropriate
key; different types of components cannot be confused.  Including redundancy
in message bodies can satisfy these assumptions.  

Some published attacks involve accepting a nonce as a key~\cite{lowe-new} or
regarding one component as being two~\cite{clark-security}.  One could alter
the model to admit type confusion attacks, but a little explicitness in
protocols~\cite{abadi-prudent} can cheaply prevent them.

\subsection{The Operators $\parts$, $\analz$ and $\synth$}\label{sec:operators}

Three operations are defined on possibly infinite sets of messages.  Each is
defined inductively, as the least set closed under specified extensions.  Each
extends a set of messages~$H$ with other items derivable from~$H$.
Typically, $H$ contains an agent's initial knowledge and the history of all
messages sent in a trace.

The set $\parts H$ is obtained from~$H$ by repeatedly adding the components of
compound messages and the bodies of encrypted messages.  (It does not regard
the key~$K$ as part of~$\Crypt K X$ unless $K$ is part of~$X$ itself.)  It
represents the set of all components of~$H$ that are potentially recoverable,
perhaps using additional keys.  Proving $X\not\in\parts H$ establishes that
$X$ does not occur in~$H$ (except, possibly, in hashed form).  Here are two
facts proved about $\parts$:
\begin{align*}
    \Crypt K X\in \parts H  &\Imp X\in\parts H\\
    \parts G \un \parts H &= \parts(G \un H).
\end{align*}

The set $\analz H$ is obtained from~$H$ by repeatedly adding the components of
compound messages and by decrypting messages whose keys are in $\analz H$. The
set represents the most that could be gleaned from~$H$ without breaking
ciphers.  If
$K\not\in\analz H$, then nobody can learn~$K$ by listening to~$H$.  Here are
some facts proved about $\analz$:
\begin{align*}
    \Crypt K X\in \analz H,\; K^{-1}\in\analz H  &\Imp X\in\analz H\\
    \analz G \un \analz H &\sbs \analz(G \un H)\\
    \analz H &\sbs \parts H.
\end{align*}

The set $\synth H$ models the messages a spy could build up from elements
of~$H$ by repeatedly adding agent names, forming compound messages and
encrypting with keys contained in~$H$.  Agent names are added because they are
publicly known.  Nonces and keys are not added because they are unguessable;
the spy can only use nonces and keys given in~$H$.  Here are two facts proved
about $\synth$:
\begin{align*}
    X\in \synth H,\; K\in H  &\Imp \Crypt K X\in\synth H\\
    K\in \synth H&\Imp K\in H.
\end{align*}

\subsection{The Attacker}

The enemy observes all traffic in the network---the set $H$---and sends
fraudulent messages drawn from the set $\synth(\analz H)$.  Interception of
messages is modelled indirectly: any message can be ignored.

No protocol should demand perfect competence from all players.  If the spy
should get hold of somebody's key, communications between other agents should
not suffer.  The model gives the spy control over an unspecified set of
compromised agents; he holds their private keys.  Most protocol descriptions
include an Oops event to allow accidental loss of session keys.

Our spy is accepted by the others as an honest agent.  He may send normal
protocol messages using his own long-term secret key, as well as sending
fraudulent messages.  This combination lets him participate in protocol runs
using intercepted keys, thereby impersonating other agents.

The spy is powerful, but he is the same in all protocols.  A common body of
laws and \emph{tactics} (mechanical proof procedures) is available.  A tactic
often proves the spy's case of the induction automatically.

\subsection{Modelling a Protocol}\label{sec:modelling}

Most events in a trace have the form $\Says{A}{B}{X}$, which means `$A$ sends
message~$X$ to~$B$.'  Another possible event is $\Notes{A}{X}$, which means
`$A$ stores $X$ internally.'  Other events could be envisaged, such as the
replacement of a long-term key.  Each agent's state is represented by its
initial knowledge (typically, its private key) and what it can
scan from the list of events.  Apart from the spy, agents only read messages
addressed to themselves.  The event $\Notes{A}{X}$ is visible to~$A$ and, if
$A$ is compromised, to the spy.

Consider a variant of the Otway-Rees protocol~\cite[page
247]{ban89}:
\begin{alignat*}{2}
  &1.&\;  A\to B  &: \Na, A, B, \comp{\Na,A,B}_{\Ka} \\
  &2.&\;  B\to S  &: \Na, A, B, \comp{\Na,A,B}_{\Ka}, \Nb, \comp{\Na,A,B}_{\Kb} \\
  &3.&\;  S\to B  &: \Na, \comp{\Na,\Kab}_{\Ka}, \comp{\Nb,\Kab}_{\Kb} \\
  &4.&\;  B\to A  &: \Na, \comp{\Na,\Kab}_{\Ka}
\end{alignat*}

Informally, (1) $A$ contacts~$B$, generating~$\Na$ to identify the run.  Then
(2) $B$ forwards $A$'s message to the authentication server, adding a nonce of
his own.  Then (3) $S$ generates a new session key~$\Kab$ and packages it
separately for~$A$ and~$B$.  Finally, (4) $B$ decrypts his part of message~3,
checks that the nonce is that sent previously, and forwards the rest to~$A$,
who will similarly compare nonces before accepting $\Kab$.

The protocol steps are modelled as possible extensions of a trace with new
events.  The server is the constant $\Server$, while $A$ and $B$ are variables
ranging over all agents, including $\Server$ and the spy.  We transcribe each
step in turn:
\begin{enumerate}
\item If $evs$ is a trace, $\Na$ is a fresh nonce and $B$ is an agent distinct
from~$A$ and~$\Server$, then $evs$ may be extended with the event 
\[ \Says{A}{B}{\comp{\Na,A,B, \comp{\Na,A,B}_{\Ka}}}. \]

\item If $evs$ is a trace that has an event of the form
\[ \Says{A'}{B}{\comp{\Na,A,B,X}}, \]
and $\Nb$ is a fresh nonce and
  $B\not=\Server$, then $evs$ may be extended with the event
\[ \Says{B}{\Server}{\comp{\Na,A,B,X,\Nb,\comp{\Na,A,B}_{\Kb}}}. \]
The sender's name is shown as~$A'$ and is not used in the new event because~$B$
cannot know who really sent the message.  The component intended to be
encrypted with $A$'s key is shown as~$X$, because~$B$ does not attempt to read
it.

\item If $evs$ is a trace containing an event of the form
\begin{align*}
\Says{B'}{\Server}{\comp{\Na,A,B,&\comp{\Na,A,B}_{\Ka},\Nb,\\
                                 &\comp{\Na,A,B}_{\Kb}}} 
\end{align*}
and $\Kab$ is a fresh key and $B\not=\Server$, then
$evs$ may be extended with the event 
\[ \Says{\Server}{B}{\comp{\Na,\comp{\Na,\Kab}_{\Ka},\comp{\Nb,\Kab}_{\Kb}}}. \]
The server too does not know where the message originated, hence the~$B'$
above.  If he can decrypt the components using the keys of the named agents,
revealing items of the right form, then he accepts the message as valid and
replies to~$B$.

\item If $evs$ is a trace containing the two events
\begin{align*}
 \SAYS{B}\,{\Server} & \,{\comp{\Na,A,B,X',\Nb,\comp{\Na,A,B}_{\Kb}}} \\
 \SAYS{S'}\,{B}      & \,{\comp{\Na,X,\comp{\Nb,K}_{\Kb}}} 
\end{align*}
and $A\not=B$, then
$evs$ may be extended with the event 
\[ \Says{B}{A}{\comp{\Na,X}}. \]
Agent $B$ receives a message of the expected format, decrypts his portion,
checks that $\Nb$ agrees with the nonce he previously sent to the server, and
forwards component~$X$ to~$A$.  The sender of the first message is shown
as~$B$ because $B$ knows if he has sent such a message.  The rule does not
specify the message from~$S'$ to be more recent than that from~$B$; this holds
by the freshness of~$\Nb$.
\end{enumerate}

There is a fifth, implicit, step, in which $A$ checks her nonce and confirms
the session.  Implicit steps can be modelled, if necessary.  For Otway-Rees,
it suffices to prove authenticity of the certificate that $A$ receives in
step~4.  For TLS~\cite{tls-1.0,paulson-tls}, the model includes a rule for
session confirmation in order to support the resumption of past sessions.

We cannot assume that a message sent in step~$i$ will be received.  But we can
identify the sending of a message in step~$i+1$ with the receipt of a
satisfactory message in step~$i$.  Because the model never forces agents to
act, there will be traces in which $A$ sends~$X$ to~$B$ but $B$ never
responds.  We may interpret such traces as indicating that $X$ was
intercepted, $B$ rejected~$X$, or $B$ was down.

An agent may participate in several protocol runs concurrently; the trace
represents his state in all those runs.  He may respond to past events, no
matter how old they are.  He may respond any number of times, or never.  If
the protocol is safe even under these liberal conditions, then it will remain
safe when time-outs and other checks are added.  Letting agents respond only
to the most recent message would prevent modelling middle-person attacks.
Excluding some traces as ill-formed weakens theorems proved about all traces.

\subsection{Standard Rules}\label{sec:oops}

A protocol description usually requires three additional rules.  One is
obvious: the empty list, [], is a trace.  Two other rules model fake messages
and accidents.

If $evs$ is a trace, $X\in\synth(\analz H)$ is a fraudulent message and
$B\not=\Spy$, then $evs$ may be extended with the event 
\[ \Says{\Spy}{B}{X}. \]
Here $H$ contains all messages in the past trace.  It includes the spy's
initial state, which holds the long-term keys of an arbitrary set of `bad'
agents.  The spy may say anything he plausibly could say and can masquerade as
any of the bad agents.

The TLS protocol~\cite{tls-1.0} arrives at session keys by exchanging nonces
and applying a pseudo-random-number function.  I have modelled
TLS~\cite{paulson-tls} by assuming this function to be an arbitrary
injection.  In the protocol specification, agents apply the random-number
function when necessary.  The spy has an additional rule that allows him to
apply the function to any message items at his disposal.  Other protocols in
which keys are computed will require an analogous rule.

If $evs$ is a trace and $\Server$ distributed the session key~$K$ in a run
involving the nonces $\Na$ and~$\Nb$, then $evs$ may be extended with the event
\[ \Notes{\Spy}{\comp{\Na,\Nb,K}}. \]
This strange-looking rule, the Oops rule, models the loss (by any means) of
session keys.  We need an assurance that lost keys cannot compromise future
runs.  The Oops message includes nonces in order to identify the protocol run,
distinguishing between recent and past losses.

For some protocols, such as Yahalom, the Oops rule brings hidden properties to
light~\cite{paulson-yahalom}.  For others, it is not clear whether Oops can be
expressed at all.

\subsection{Induction}\label{sec:induction}

The specification defines the set of possible traces {\em inductively}: it is
the least set closed under the given rules.  To appreciate what this means, it
may be helpful to recall that the set~$\N$ of natural numbers is inductively
defined by the rules $0\in\N$ and $n\in\N\Imp\Suc n\in\N$.

For reasoning about an inductively defined set, we may use the corresponding
induction principle.  For the set $\N$, it is the usual mathematical
induction: to prove~$P(n)$ for each natural number $n$, prove~$P(0)$ and prove
$P(x)\Imp P(\Suc x)$ for each $x\in \N$.  For the set of traces, the induction
principle says that $P(evs)$ holds for each trace~$evs$ provided $P$ is
preserved under all the rules for creating traces.

We must prove $P[]$ to cover the empty trace.  For each of the other rules,
we must prove an assertion of the form $P(evs)\Imp P(ev\#evs)$, where event
$ev$ contains the new message.  (Here $ev\#evs$ is the trace that
extends~$evs$ with event~$ev$: new events are added to the front of a
trace.)  The rule may resemble list induction, but the latter considers all
conceivable messages, not just those allowed by the protocol.

A trivial example of induction is to prove that no agent sends a
message to himself: no trace contains an event of the form $\Says{A}{A}{X}$. 
This holds vacuously for the empty trace, and the other rules specify
conditions such as $B\not=\Server$ to prevent the creation of such events.

\subsection{Regularity Lemmas}

These lemmas concern occurrences of a particular item $X$ as a possible
message component.  Such theorems have the form $X\in\parts H\imp\cdots$,
where $H$ is the set of all messages available to the spy.  These are strong
results: they hold in spite of anything that the spy might do.

For most protocols, it is easy to prove that the spy never gets hold of any
agent's long-term key, excluding the bad agents.  The inductive proof amounts
to examining the protocol rules and observing that none of them involve
sending long-term keys.  The spy cannot send any either because, by the
induction hypothesis, he has none at his disposal except those of the bad
agents.

Unicity results state that nonces or session keys identify certain messages.
Naturally we expect the server never to re-issue session keys, or agents their
nonces.  If they choose these items to be fresh, then it is straightforward to
prove that the key (or nonce) part of a message determines the values of the
other parts.

\subsection{Secrecy Theorems}\label{sec:secrecy}

Regularity lemmas are easy to prove because they are stated in terms of the
$\parts$ operator.  Secrecy cannot be so expressed; if $X$ is a secret then
some agents can see~$X$ and others cannot. Secrecy theorems are, instead,
stated in terms of $\analz$.  Their proofs can be long and difficult, typically
splitting into cases on whether or not certain keys are compromised.

A typical result involving $\analz$ states that if the
spy holds some session keys, he cannot use them to reveal others.  It would
suffice to prove that nobody sends messages of the form $\Crypt
\Kab\,\comp{\ldots \Kcd \ldots}$, but this claim is false: the spy can send such
messages and make other agents send them.  Fortunately, he
does not thereby learn new session keys; to work such mischief, he must
already possess~$\Kcd$.

The discussion above suggests the precise form of the theorem.  If $K$ can be
obtained with the help of a session key~$K'$ and previous traffic,
then either $K=K'$ or $K$ can be obtained from the traffic alone.
Because some protocol steps introduce new keys, proof by induction seems to
require strengthening the formula, generalizing $K'$ to a set of session keys.
This is the \emph{session key compromise theorem}.

Proving a theorem of this form is often the hardest task in analyzing a
protocol.  A huge case analysis often results.  While it can be automated, the
processor time required seems to be exponential in the number of different
keys used for encryption in any single protocol message.  A bit of creativity
here can yield substantial savings; see \S\ref{sec:coarser} below.  For simple
key-exchange protocols, however, essentially the same six-command proof script
always seems to work.

The theorem makes explicit something we may have taken for granted: that no
agent should use session keys to encrypt other keys (see also Gollmann
\cite[\S2.1]{gollmann-what}).  A generalization of the theorem can be used to
prove the secrecy of $B$'s nonce in Yahalom~\cite{paulson-yahalom}.

The \emph{session key secrecy theorem} states that if the server distributes a
session key $\Kab$ to~$A$ and~$B$, then the spy never gets this key.  Since the
spy is treated in every respect as an honest agent, we may conclude that no
other agent gets the key either, even by accident.

The theorem stipulates that $A$ and~$B$ are uncompromised and that no
Oops message has given the session key to the spy.  If we must forbid all Oops
messages for~$\Kab$, not just those involving the current nonces, then we should
consider whether the protocol is vulnerable to a replay attack.

Secrecy properties can usually be proved using six or seven commands.  A
constant problem in secrecy proofs is being presented with gigantic formulas.
We need to discard just the right amount of information and think carefully
about how induction formulas are expressed.

\subsection{Finding Attacks}\label{sec:attack}

Secrecy is necessary but not sufficient for correctness.  The server might be
distributing the key to the wrong pair of agents.  When $A$ receives message~4
of the Otway-Rees protocol, can she be sure it really came from~$B$, who got it
from~$S$?  For the simplified version of the protocol outlined above
(\S\ref{sec:modelling}), the answer is no.  

The only secure part of message~4 is its encrypted part,
$\comp{\Na,\Kab}_{\Ka}$.  But it need not have originated as the first
encrypted part of message~3.  It could as well have originated as the second
part, if $S$ received a fraudulent message~2 in which a previous~$\Na$ had
been substituted for~$\Nb$.

The machine proof leads us to consider a scenario in which $\Na$ is used in
two roles.  It is then easy to invent an attack.  A spy, $C$, intercepts $A$'s
message~1 and records~$\Na$.  He masquerades first as~$A$ (indicated as $C_A$
below), causing the server to issue him a session key~$\Kca$ and also to
package $\Na$ with this key.  He then masquerades as~$B$.
\begin{alignat*}{2}
  &1.&\;   A\to C_B  &: \Na, A, B, \comp{\Na,A,B}_{\Ka} \\
  &1'.&\;  C\to A  &: \Nc, C, A, \comp{\Nc,C,A}_{\Kc}
  \displaybreak[0] \\
  &2'.&\;  A\to C_S  &: \Nc, C, A, \comp{\Nc,C,A}_{\Kc},
                        \Na', \comp{\Nc,C,A}_{\Ka}\\
  &2''.&\; C_A\to S  &: \Nc, C, A, \comp{\Nc,C,A}_{\Kc}, \Na,
                                 \comp{\Nc,C,A}_{\Ka}
  \displaybreak[0] \\
  &3'.&\;  S\to C_A  &: \Nc, \comp{\Nc,\Kca}_{\Kc}, \comp{\Na,\Kca}_{\Ka}  \\
  &4.&\;  C_B\to A  &: \Na, \comp{\Na,\Kca}_{\Ka}
\end{alignat*}
Replacing nonce~$\Na'$ by~$\Na$ in message~$2'$ eventually causes~$A$ to
accept key~$K_{ca}$ as a key for talking with~$B$, because $\Na$ is $A$'s
original nonce.  This attack is more serious than that discovered by Mao and
Boyd~\cite{mao-formal}, where the server could detect the repetition of a
nonce.  It cannot occur in the original version of Otway-Rees, where $\Nb$ is
encrypted in the second message.

Otway-Rees uses nonces not just to assure freshness, but for binding: to
identify the principals~\cite{abadi-prudent}.  Verifying the binding
complicates the formal proofs.  One can prove---for the corrected
protocol---that $\Na$ and $\Nb$ uniquely identify the messages they originate
in and never coincide.  Then we can prove guarantees for both agents: if they
receive the expected messages, and the nonces agree, then the server really
did distribute the session key to the intended parties.

\section{A Mechanized Theory of Messages}\label{sec:theory}

The approach has been mechanized using Isabelle/HOL, an instantiation of the
generic theorem prover Isabelle~\cite{paulson-isa-book,paulson-markt} to
higher-order logic.  Isabelle is appropriate because of its support for
inductively defined sets and its automatic tools.  Some
Isabelle syntax appears below in order to convey a feel for how proofs are
conducted.

The methodology is tailored to Isabelle and makes heavy use of its classical
reasoner~\cite{paulson-generic}.  However, it can probably be modified to suit
other higher-order logic provers such as PVS~\cite{pvs-fault} or
HOL~\cite{mgordon-hol}.  At a minimum, the prover should provide a simplifier
that takes conditional rewrite rules and that can perform automatic case
splits for if-then-else expressions.  Unless some form of set theory is
available, the algebraic laws for $\parts$, $\analz$ and $\synth$ will be
lost.  HOL predicates make satisfactory sets, but finite lists do not.

Isabelle/HOL has a polymorphic type system resembling ML's~\cite{paulson-ml2}.
An item of type $\agent$ can never appear where something of type $\msg$ is
expected.  Type inference eliminates the need to specify types in expressions.
Laws about lists, sets, etc., are polymorphic; the rewriter uses
the appropriate types automatically.

\subsection{Agents and Messages}

There are three kinds of agents: the server~$\Server$, the friendly agents,
and the spy.  Friendly agents have the form $\Friend i$, where $i$ is a
natural number.  The following declaration specifies type $\agent$ to
Isabelle.  (Note that $\Server$ is called \texttt{Server} and that $\nat$ is
the type of natural numbers.)
\begin{ttbox}
datatype agent = Server | Friend nat | Spy
\end{ttbox}
A datatype declaration creates a union type, with injections whose
ranges are disjoint.  It follows that the various kinds of agent are distinct,
with $\Server\not=\Friend i$, $\Server\not=\Spy$, $\Spy\not=\Friend i$, and
moreover $\Friend i=\Friend j$ only if $i=j$.

The various kinds of message items (discussed above, \S\ref{sec:messages}) are
declared essentially as shown below.  Observe the use of type $\agent$
and the recursive use of type $\msg$.  Not shown are further declarations that
make $\comp{X_1,\ldots X_{n-1},X_n}$ abbreviate
$\mathsf{MPair}\,X_1\,\ldots\allowbreak(\mathsf{MPair}\,X_{n-1}\,X_n)$.
\begin{ttbox}
datatype msg = Agent agent
             | Number nat       (*guessable*)
             | Nonce nat        (*non-guessable*)
\pagebreak[0]             | Key   key
             | MPair msg msg
\pagebreak[0]             | Hash  msg
             | Crypt key msg
\end{ttbox}
Again, the various kinds of message are distinct, with $\Agent A\not=\Nonce N$
and so forth.  The injections $\Agent$, $\Number$, $\Nonce$ and $\Key$ are
simply type coercions.  

Because the datatype creates injections, hashing is collision-free: we have
$\Hash X=\Hash X'$ only if $X=X'$.  Encryption is strong.  Injectivity yields
the law
\[ \Crypt K X=\Crypt K' X' \Imp K=K' \conj X=X'. \]
Moreover, the spy cannot alter an encrypted message without first decrypting
it using the relevant key.  Exclusive-or violates these assumptions, as does
RSA~\cite{rsa78} unless redundancy is incorporated.  Such forms of encryption
could be modelled, but the loss of injectiveness would complicate the theory.

\subsection{Defining $\parts$, $\analz$ and $\synth$}

The operators $\parts$, $\analz$ and $\synth$ are defined inductively, as are
protocols themselves.  If $H$ is a set of messages then $\parts H$ is the
least set including~$H$ and closed under projection and decryption.  Formally,
it is defined to be the least set closed under the following rules.
\begin{gather*}
  \infer{X\in\parts H}{X\in H} \qquad
  \infer{X\in\parts H}{\Crypt K X\in\parts H} \\[1.5ex]
  \infer{X\in\parts H}{\comp{X,Y}\in\parts H} \qquad
  \infer{Y\in\parts H}{\comp{X,Y}\in\parts H}
\end{gather*}
Similarly, $\analz H$ is defined to be the least set including~$H$ and closed under
projection and decryption by known keys.
\begin{gather*}
  \infer{X\in\analz H}{X\in H} \qquad
  \infer{X\in\analz H}{\Crypt K X\in\analz H & K^{-1}\in\analz H} \\[1.5ex]
  \infer{X\in\analz H}{\comp{X,Y}\in\analz H} \qquad\qquad\qquad\qquad 
  \infer{Y\in\analz H}{\comp{X,Y}\in\analz H}
\end{gather*}
Finally, $\synth H$ is defined to be the least set that includes~$H$, agent
names and guessable numbers, and is closed under pairing, hashing and
encryption.
\begin{gather*}
  {\Agent A\in\synth H}                  \qquad
  {\Number N\in\synth H}                 \\[1.5ex]
  \infer{X\in\synth H}{X\in H} \qquad
  \infer{\Hash X\in\synth H}{X\in H}       \\[1.5ex]
  \infer{\comp{X,Y}\in\synth H}{X\in\synth H & Y\in\synth H} \qquad 
  \infer{\Crypt K X\in\synth H}{X\in\synth H & K\in H} 
\end{gather*}
To illustrate Isabelle's syntax for such definitions, here is the one for
$\analz$. 
\begin{ttbox}
consts  analz   :: msg set => msg set
inductive "analz H"
 intrs 
   Inj  "X\(\,\in\,\)H \(\Imp\) X\(\,\in\,\)analz H"
   Fst  "\{|X,Y|\}\(\,\in\,\)analz H \(\Imp\) X \(\,\in\,\)analz H"
   Snd  "\{|X,Y|\}\(\,\in\,\)analz H \(\Imp\) Y \(\,\in\,\)analz H"
   Decrypt "[| Crypt K X \(\,\in\,\)analz H;  Key(invKey K)\(\,\in\,\)analz H |] 
            \(\Imp\) X \(\,\in\,\)analz H"
\end{ttbox}
Given such a definition, Isabelle defines an
appropriate fixedpoint and proves the desired rules.  These include the
introduction rules (those that constitute the definition itself) as well as
case analysis and induction.

\label{sec:hashing}

The definition of $\parts$ does not make~$X$ a part of $\Hash X$ even
though it is a part of $\Crypt K X$.  There is no inconsistency here: for
typical protocols, private keys might be included in hashes (where they serve
as signatures) but never in encrypted messages.  We can prove that
uncompromised private keys are not part of any traffic, and use this basic
lemma to prove deeper properties.

\subsection{Derived Laws Governing the Operators}\label{sec:laws}

Section~\ref{sec:operators} presented a few of the laws proved for the
operators, but protocol verification requires many more.  Let us examine them
systematically.  All have been mechanically proved from the preceding
definitions.

The operators are monotonic: if $G\sbs H$ then
\[\parts G\sbs \parts H \quad  \analz G \sbs \analz H  \quad  
  \synth G\sbs \synth H. \]
They are idempotent:
\begin{align*}
  \parts (\parts H) &= \parts H \\
  \analz (\analz H) &= \analz H \\ 
  \synth (\synth H) &= \synth H. 
\end{align*}
Similarly, we have the equations
\[\parts (\analz H) = \parts H  \quad \analz (\parts H) = \parts H. \]
Building up, then breaking down, results in two less trivial equations:
\begin{align*}
   \parts (\synth H) &= \parts H \un \synth H \\
   \analz (\synth H) &= \analz H \un \synth H 
\end{align*}
We have now considered seven of the nine possible combinations involving two of
the three operators.  The remaining combinations, $\synth (\parts H)$ and 
$\synth (\analz H)$, appear to be irreducible.  The latter one models the
fraudulent messages that a spy could derive from~$H$.  We can still prove 
laws such as 
\[  \comp{X,Y}\in\synth (\analz H)\iff
  X\in\synth (\analz H) \conj Y\in \synth(\analz H).
\]
More generally, we can derive a bound on what the enemy can say:
\[ \infer{\parts(\INS{X}H)\sbs\synth(\analz H)\cup\parts H}
         {X\in\synth(\analz H)}
\]
$H$ is typically the set of all messages sent during a trace.  The
rule eliminates the fraudulent message~$X$, yielding an upper bound
on $\parts(\INS{X}H)$.  Typically, $\parts H$ will be bounded by an
induction hypothesis.  There is an analogous rule for $\analz$.%
\footnote{The Isabelle theories represent the set $\{X\}\un H$ by
  $\mathop{\mathsf{insert}}{X}\,H$, and similarly $\{X,Y\}\un H$ by
  $\mathop{\mathsf{insert}}{X}\,(\mathop{\mathsf{insert}}{Y}\,H)$, etc.}

\subsection{Rewrite Rules for Symbolic Evaluation}

Applying rewrite rules to a term such as 
\[ \parts\{\comp{\Agent A,\Nonce\Na}\} \]
can transform it to the equivalent three-element set
\[ \{ \comp{\Agent A,\Nonce\Na}, \Agent A, \Nonce\Na \}. \]
This form of evaluation can deal with partially specified arguments such as
$\{\comp{\Agent A,X}\}$ and 
\[ \INS{\comp{\Agent A,\Nonce\Na}}H. \]

Symbolic evaluation for $\parts$ is straightforward.  For a protocol step
that sends the message~$X$ we typically consider a subgoal containing the
expression $\parts(\INS{X}H)$ or $\analz(\INS{X}H)$.  The previous section
has discussed the case in which $X$ is fraudulent.  In other cases, $X$ will
be something more specific, such as
\begin{align*}
    \comp{&\Nonce \Na, \Agent A, \Agent B,\\
          &\Crypt \Ka \,\comp{\Nonce \Na, \Agent A, \Agent B}}.
\end{align*}
Now $\parts(\INS{X}H)$ expands to a big expression involving all the new
elements that are inserted into the set $\parts H$, from $\Nonce \Na$ and
$\Agent A$ to $X$ itself.  The expansion may sound impractical, but a subgoal
such as $\Key K\not\in\parts(\INS{X}H)$ simplifies to $\Key K\not\in\parts H$
(for the particular $X$ shown above) because none of the new elements has the
form $\Key K'$.  If this element were present, then the subgoal would still
simplify to a manageable formula, $K\not=K'\conj\Key K\not\in\parts H$.

The rules for symbolic evaluation of $\parts$ are fairly obvious.  They have
straightforward inductive proofs.
\begin{align*}
    \parts \emptyset &= \emptyset \\                                  
    \parts (\INS{\Agent A}\,H) & 
          = \INS{\Agent A} \parts H \\
    \parts (\INS{\Nonce N}\,H) & 
          = \INS{\Nonce N} \parts H \\
    \parts (\INS{\Key K}\,H) & 
          = \INS{\Key K} \parts H \\
    \parts (\INS{\comp{X,Y}}H) &
          = \INS{\comp{X,Y}} \parts (\INS{X}\INS{Y}H)\\
    \parts (\INS{\Hash X}\,H) & = \INS{\Hash X} \parts H \\
    \parts (\INS{\Crypt K X}H) &
          = \INS{\Crypt K X} \parts (\INS{X}H)
\end{align*}
Symbolic evaluation of $\analz$ is more difficult.  Let us first define the
set of keys that can decrypt messages in~$H$:
\[ \keysFor H \eqdef \{K^{-1}\mid \exists X.\,\Crypt K X\in H\} \]
A key can be pulled through $\analz$ if it is not needed for decryption.
\[   \infer{\analz (\INS{\Key K}\,H)  = \INS{\Key K}(\analz H)}
           {K\not\in\keysFor(\analz H)}
\]
The rewrite rule for encrypted messages involves case analysis on whether or
not the matching key is available.
\begin{multline*}
   \analz (\INS{\Crypt K X}H) = \\
     \begin{cases}
       \INS{\Crypt K X}(\analz (\INS{X}H)) & K^{-1}\in \analz H \\
       \INS{\Crypt K X}(\analz H)          & \text{otherwise}
     \end{cases}
\end{multline*}
Nested encryptions give rise to nested if-then-else expressions.  Sometimes we
know whether the relevant key is secure, but letting automatic tools
generate a full case analysis gives us short proof scripts.  Impossible cases
are removed quickly.  Redundant case analyses---those that simplify to `if
$P$ then $Q$ else $Q$'---can be simplified to~$Q$.  The resulting expression
might be enormous, but symbolic evaluation at least expresses $\analz
(\INS{X}H)$ in terms of $\analz H$, which should let us invoke the induction
hypothesis.

Rewriting by the following rule, which is related to idempotence, simplifies
the cases that arise when an agent forwards to another agent some message
that is visible in previous traffic.
\[ \infer{\analz (\INS{X}H) = \analz H}{X\in\analz H} \]

Symbolic evaluation of $\synth$ is obviously impossible: its result is
infinite.  Fortunately, it is never necessary.  Instead, we need to simplify
assumptions of the form $X\in\synth H$, which arise when considering whether a
certain message might be fraudulent.  The inductive definition regards nonces
and keys as unguessable, giving rise to the implications
\begin{align*}
    \Nonce N\in\synth H &\Imp \Nonce N\in H\\
    \Key K\in\synth H &\Imp \Key K\in H
\end{align*}
If $\Crypt K X\in\synth H$ then either $\Crypt K X\in H$ or else $X\in\synth
H$ and $K\in H$.  If we already know $K\not\in H$, then the rule tells us that
the encrypted message is a replay rather than a forgery.  There are similar
rules for $\Hash X\in\synth H$ and $\comp{X,Y}\in\synth H$.

The facts mentioned in this section are among over 110 theorems that have been
proved about $\parts$, $\analz$, $\synth$ and $\keysFor$.  Most of them are
stored in such a way that Isabelle can apply them automatically for
simplification.  Logically speaking, some of these proofs are complex.  They
need on average under two commands (tactic invocations) each, thanks to
Isabelle's automatic tools.  The full proof script, over 210 commands,
executes in under 45 seconds.

\subsection{Events and Intruder Knowledge}\label{sec:events}

A trace is a list of events, each of the form $\Says{A}{B}{X}$ or
$\Notes{A}{X}$.  Isabelle/HOL provides lists, while events are trivial to
declare as a datatype.
\begin{ttbox}
datatype event = Says  agent agent msg
               | Notes agent       msg
\end{ttbox}

Otway-Rees assumes a symmetric-key environment.  Every agent~$A$ has a
long-term key, $\shrK A$, shared with the server.  The spy has such a key
($\shrK\Spy$) and there is even the redundant $\shrK\Server$.  Function
$\initState$ specifies agents' initial knowledge.  The spy knows the
long-term keys of the agents in the set $\bad$.
\begin{align*}
    \initState\Server & \eqdef \text{all long-term keys}\\
    \initState(\Friend i) & \eqdef \{\Key(\shrK(\Friend i))\}\\
    \initState\Spy & \eqdef \{\Key(\shrK(A))\mid A\in\bad\}
\end{align*}

The function $\mathsf{spies}$ models the set of messages the spy can see in a
trace.  He sees all messages sent across the network.  He even sees the
internal notes of the bad agents, who can be regarded as being under his
control.  From the empty trace, he sees only his initial state.  Recall that
$ev\#evs$ is the list consisting of~$ev$ prefixed to the list~$evs$.
\begin{align*}
   \spies{[]} &\eqdef \initState\Spy\\[1ex]
   \spies{((\Says{A}{B}{X})\cons evs)} &\eqdef \INS{X} \spies{evs}\\[1ex]
   \spies{((\Notes{A}{X})\cons evs)}   &\eqdef 
      \begin{cases} \INS{X} \spies{evs}   &  \text{if } A\in\bad\\
                    \spies{evs}           &  \text{otherwise}
      \end{cases}
\end{align*}
The function $\mathsf{spies}$ describes the spy's view of traffic in order to
formalize message spoofing.  For other agents, the formal protocol rules
mention previous messages directly.

The set $\used evs$ formalizes the notion of freshness.  The set includes
$\parts(\spies evs)$ as well as the $\parts$ of all messages held privately by
any agent.  For example, if $\Key K\not\in\used evs$, then $K$ is fresh (in
$evs$) and differs from all long-term keys.
\begin{align*}
   \used{[]} &\eqdef \union B.\, \parts(\initState B)\\
   \used{((\Says{A}{B}{X})\cons evs)} &\eqdef \parts\{X\} \un \used{evs}\\
   \used{((\Notes{A}{X})\cons evs)}   &\eqdef \parts\{X\} \un \used{evs}
\end{align*}

\begin{figure*}[htbp]
\begin{ttbox}
\RuleLabel{Nil} [] \(\in\) otway 

\RuleLabel{Fake} [| evs \(\in\) otway;  B\(\not=\)Spy;  X \(\in\) synth (analz (spies evs)) |]
     \(\Imp\) Says Spy B X  # evs  \(\in\) otway 

\RuleLabel{OR1} [| evs1 \(\in\) otway;  A \(\not=\) B;  B \(\not=\) Server;  Nonce NA \(\not\in\) used evs1 |]
     \(\Imp\) Says A B \{|Nonce NA, Agent A, Agent B, 
                    Crypt (shrK A) \{|Nonce NA, Agent A, Agent B|\} |\} 
          # evs1  \(\in\) otway 

\RuleLabel{OR2} [| evs2 \(\in\) otway;  B \(\not=\) Server;  Nonce NB \(\not\in\) used evs2;
        Says A' B \{|Nonce NA, Agent A, Agent B, X|\} \(\in\) set evs2 |]
     \(\Imp\) Says B Server 
             \{|Nonce NA, Agent A, Agent B, X, Nonce NB, 
               Crypt (shrK B) \{|Nonce NA, Agent A, Agent B|\}|\}
          # evs2  \(\in\) otway 

\RuleLabel{OR3} [| evs3 \(\in\) otway;  B \(\not=\) Server;  Key KAB \(\not\in\) used evs3;
        Says B' Server 
             \{|Nonce NA, Agent A, Agent B, 
               Crypt (shrK A) \{|Nonce NA, Agent A, Agent B|\}, 
               Nonce NB, 
               Crypt (shrK B) \{|Nonce NA, Agent A, Agent B|\}|\}
          \(\in\) set evs3 |]
     \(\Imp\) Says Server B 
             \{|Nonce NA, 
               Crypt (shrK A) \{|Nonce NA, Key KAB|\},
               Crypt (shrK B) \{|Nonce NB, Key KAB|\}|\}
          # evs3  \(\in\) otway 

\RuleLabel{OR4} [| evs4 \(\in\) otway;  A \(\not=\) B;  
        Says B Server \{|Nonce NA, Agent A, Agent B, X', Nonce NB,
                        Crypt (shrK B) \{|Nonce{\ts}NA, Agent{\ts}A, Agent{\ts}B|\}|\}
          \(\in\) set evs4;
        Says S' B \{|Nonce NA, X, Crypt (shrK B) \{|Nonce NB, Key K|\}|\}
          \(\in\) set evs4 |]
     \(\Imp\) Says B A \{|Nonce NA, X|\} # evs4  \(\in\) otway 

\RuleLabel{Oops} [| evso \(\in\) otway;  B \(\not=\) Spy;
        Says Server B \{|Nonce NA,{\ts}X,{\ts}Crypt\ts(shrK B)\ts\{|Nonce NB,{\ts}Key K|\}|\}
          \(\in\) set evso |]
     \(\Imp\) Notes Spy \{|Nonce NA, Nonce NB, Key K|\} # evso  \(\in\) otway 
\end{ttbox}
\caption{Specifying the Otway-Rees Protocol} \label{fig:otway}
\end{figure*}

\section{A Shared-Key Protocol: Otway-Rees}\label{sec:otway}

Section~\ref{sec:modelling} discussed the modelling of a protocol informally,
though in detail.  Now, let us consider the specification supplied to the
theorem prover (Fig.\ts\ref{fig:otway}).  

The identifiers at the far left name the rules: Nil for the empty trace, Fake
for fraudulent messages, OR1--4 for protocol steps, and Oops for the
accidental loss of a session key.  The set of traces is the constant $\otway$.

The Nil rule is trivial, so let us examine Fake.  The condition $evs\in\otway$
states that $evs$ is an existing trace.  Now
\[ X\in \synth(\analz(\seespy)) \]
denotes any message that could be forged from what the spy could decrypt from
the trace; recall that he holds the bad agents' private keys.  The spy can
send forged messages to any other agent~$B$, including the server.  All rules
have additional conditions, here $B\not=\Spy$, to ensure that agents send no
messages to themselves; this trivial fact eliminates some impossible cases in
proofs.

Rule OR1 formalizes step~1 of Otway-Rees.  List $evs1$ is the current trace.
(Calling it $evs1$ instead of simply~$evs$ tells the user which subgoals have
arisen from this rule during an inductive proof, even after case-splitting,
etc.)  The nonce~$\Na$ must be fresh: not contained in $\used evs1$.
An agent has no sure means of generating fresh nonces, but can do so with a
high probability by choosing enough random bytes.

In rule OR2, $\sol evs2$ denotes the set of all events, stripped of their
temporal order.  Agent $B$ responds to a past message, no matter how old it
is.  We could restrict the rule to ensure that $B$ never responds to a given
message more than once.  Current proofs do not require this
restriction, however, and it might prevent the detection of replay attacks.

There is nothing else in the rules that was not already discussed
in~\S\ref{sec:modelling} above.  Translating informal protocol notation into
Isabelle format is perhaps sufficiently straightforward to be automated.

\subsection{Proving Possibility Properties}

The first theorems to prove of any protocol description are some
\emph{possibility properties}.  These do not assure liveness, merely that
message formats agree from one step to the next.  We cannot prove that
anything must happen; agents are never forced to act.  But if the protocol can
never proceed from the first message to the last, then it must have been
transcribed incorrectly.

Here is a possibility property for Otway-Rees.  For all agents $A$ and~$B$,
distinct from themselves and from the server, there is a key~$K$, nonce~$N$
and a trace such that the final message $B\to A:\Na,\comp{\Na,\Kab}_{\Ka}$ is
sent.  This theorem is proved by joining up the protocol rules in order and
showing that all their preconditions can be met.

\subsection{Proving Forwarding Lemmas}\label{sec:forwarding}

Some results are proved for reasoning about 
steps in which an agent forwards an unknown item.  Here is a rule for OR2:
\[ \infer{X\in\analz(\seespy)}
         {\Says{A'}{B}{\comp{N, \Agent A, \Agent B, X}\in\sol evs}}
\]
The proof is trivial.  The spy sees the whole of the message; since $X$ is
transmitted in clear, $\analz$ will find it.  The spy can learn nothing new
by seeing~$X$ again when $B$ responds to this message.

Sometimes the forwarding party removes a layer of encryption, perhaps
revealing something to the spy.  Then the forwarding lemma is weaker: it is
stated using $\parts$ instead of $\analz$, and is useful only for those
theorems (`regularity lemmas') that can be stated using $\parts$.  Otway-Rees
has no nested encryption, but the Oops rule removes a layer of encryption: it
takes $K$ from the server's message and gives it to the spy.  Its forwarding
lemma states that this act does not add new keys to $\parts(\seespy)$.
\[ \infer{K\in\parts(\seespy)}
         {\Says{\Server}{B}{\comp{\Na,X,\Crypt\,K'\,\comp{\Nb,K}}}\in\sol evs}
\]

\subsection{Proving Regularity Lemmas}\label{sec:regularity}

Statements of the form
$X\in\parts(\seespy) \imp\cdots$
impose conditions on the appearance of~$X$ in any message.  Many such
lemmas can be proved in the same way.  

\begin{enumerate}
\item Apply induction, generating cases for each protocol step and Nil, Fake,
  Oops.

\item For each step that forwards part of a message, apply the corresponding
  forwarding lemma, using $\analz H\sbs\parts H$ if needed to express the
  conclusion in terms of $\parts$.

\item Prove the trivial Nil case using a standard automatic tactic. 

\item Simplify all remaining cases.
\end{enumerate}
In Isabelle (or any programmable tool), the user can define a tactic to
perform these tasks and return any remaining subgoals.  Usually, the Fake
case can then be proved automatically.

A basic regularity law states that secret keys remain secret.  If 
$evs\in\otway$ (meaning, $evs$ is a trace) then
\[ \Key(\shrK A)\in\parts(\seespy)\iff A\in\bad. \]
Two commands generate the proof.

\subsection{Proving Unicity Theorems}\label{sec:unique}

Fresh session keys and nonces uniquely identify their message of origin.  But
we must exclude the possibility of spoof messages, and this can be done in two
different ways.  In the case of session keys, a typical formulation
refers to an event and names the server as the sender (for $evs\in\otway$):
\begin{multline*}
\exists B'\;\Na'\;\Nb'\;X'.\quad \forall B\;\Na\;\Nb\;X. \\
    \Says{\Server}{B}{\comp{\Na,X,\Crypt(\shrK B)\comp{\Nb,K}}\in\sol evs}\\
        \imp B=B'\conj \Na=\Na' \conj \Nb=\Nb' \conj X=X'.
\end{multline*}
The free occurrence of~$K$ in the event uniquely determines the other four
components shown.  To apply such a theorem requires
proof that the message in question really originated with the server.

An alternative formulation, here for nonces, presumes the existence of a
message encrypted with a secure key:
\begin{multline*}
\exists B'.\, \forall B. \Crypt(\shrK A)\comp{\Na,\Agent A,\Agent B}
\in\parts(\seespy) \\
\imp B=B'.
\end{multline*}
Here $evs$ is some trace and, crucially, $A\not\in\bad$.  The spy could not
have performed the encryption because he lacks $A$'s key.  The free occurrence
of~$\Na$ in the message determines the identity of~$B$.

As in the BAN logic, we obtain guarantees from encryption by keys known to
be secret.  However, such guarantees are not built into the logic: they are
proved.  Both formulations of unicity may be regarded as regularity lemmas.
Their proofs are not hard to generate.

\subsection{Proving Secrecy Theorems}\label{sec:proving-secrecy}

Section~\ref{sec:secrecy} discussed the session key compromise theorem.  If
$K$ can be obtained from a set of session keys and messages, then either it is
one of those keys, or it can be obtained from the messages alone.  The theorem
is formulated as follows, for an arbitrary trace $evs$ ($evs\in\otway$).
\[  K\in\analz({\cal K}\cup \seespy)) \iff
    K\in {\cal K} \disj K\in\analz(\seespy)
\]
Here ${\cal K}$ is an arbitrary set of session keys, not necessarily present
in the trace.  The right hand side of the equivalence is a simplification of
\[ K\in{\cal K}\cup \analz(\seespy) \]
Replacing $\analz$ by $\parts$, which distributes over union, would render the
theorem trivial.  The right-to-left direction is trivial anyway.

To prove such a theorem can be a daunting task.  However, there are techniques
that make proving secrecy theorems almost routine.
\begin{enumerate}
\item Apply induction.  

\item For each step that forwards part of a message, apply the corresponding
forwarding lemma, if its conclusion is expressed in terms of $\analz$.

\item Simplify all cases, using rewrite rules to evaluate $\analz$
  symbolically: pulling out agent names, nonces and compound messages and
  performing automatic case splits on encrypted messages.
\end{enumerate}

The Fake case usually survives, but it can be proved by a standard argument
involving the properties of $\synth$ and $\analz$.  This argument can be
programmed as a tactic, which works for all protocols investigated.  For the
session key compromise theorem, no further effort is needed.  Other secrecy
theorems require a detailed argument.  Chief among these is proving that
nonce~$\Nb$ of the Yahalom protocol~\cite{ban89} remains secret, which
requires establishing a correspondence between nonces and
keys~\cite{paulson-yahalom}.

\subsection{Proving the Session Key Secrecy Theorem}

This theorem states that the protocol is correct from the server's viewpoint.
Let $evs\in\otway$ and $A$, $B\not\in\bad$.  Suppose that the server issues
key~$K$ to $A$ and~$B$:
\begin{align*}
   \Says{\Server}{B}{\comp{\Na, & \Crypt (\shrK A) \comp{\Na, K},\\
                                & \Crypt (\shrK B) \comp{\Nb, K}}} \in\sol evs
\end{align*}
Suppose also that the key is not lost in an Oops event involving the same
nonces: 
\[ \Notes{\Spy}{\comp{\Na, \Nb, K}} \not\in\sol evs \]
Then we have $K \not\in\analz (\seespy)$; the key is never available to the
spy.

This secrecy theorem is slightly harder to prove than the previous one.  In
the step~3 case, there are two possibilities.  If the new message is the very
one mentioned in the theorem statement then the session key is not fresh,
contradiction; otherwise, the induction hypothesis yields the needed result.
Isabelle can prove the step~3 case automatically.  The Oops case is also
nontrivial; showing that any Oops message involving~$K$ must also
involve~$\Na$ and~$\Nb$ requires unicity of session keys, a theorem discussed
in the previous section.  The full proof script consists of seven commands and
executes in eight seconds, generating a proof of over 4000 steps.%
\footnote{All runtimes were measured on a 300MHz Pentium II\@.  A human
  could probably generate a much shorter proof by omitting irrelevant steps.}

\subsection{Proving Authenticity Guarantees}\label{sec:further}

The session key secrecy theorem described above is worthless on its own.  It
holds of a protocol variant that can be attacked~(\S\ref{sec:attack}).  In the
correct protocol, if $A$ or~$B$ receive the expected nonce, then the server
has sent message~3 in precisely the right form.  Agents need guarantees
(subject to conditions they can check) confirming that their certificates
are authentic.  Proving such guarantees for $A$ and~$B$ completes the
security argument, via an appeal to the session key secrecy theorem.

The correct protocol differs in message~2, which now
encrypts~$\Nb$: 
\begin{alignat*}{2}
  &1.&\; A\to B &: \Na, A, B, \comp{\Na,A,B}_{\Ka} \\
  &2.&\; B\to S &: \Na, A, B, \comp{\Na,A,B}_{\Ka}, \comp{\Na,\Nb,A,B}_{\Kb} \\
  &3.&\; S\to B &: \Na, \comp{\Na,\Kab}_{\Ka}, \comp{\Nb,\Kab}_{\Kb} \\
  &4.&\; B\to A &: \Na, \comp{\Na,\Kab}_{\Ka}
\end{alignat*}
After receiving the step~3 message, $B$ can inspect the certificate that is
encrypted with his key, but not the one he forwards to~$A$.

$B$'s guarantee states that if a trace contains an event of the form
\[ \Says{S'}{B}{\comp{\Na, X, \Crypt (\shrK B) \comp{\Nb, \Key K}}} \]
and if $B$ is uncompromised and has previously sent message~2, 
\begin{align*}
   \Says{B}{\Server}{\comp{ &\Na, \Agent A, \Agent B, X', \\
                 &\Crypt (\shrK B) \comp{\Na,\Nb, \Agent A, \Agent B}}}
\end{align*}
then the server has sent a correct instance of step~3.  The theorem does not
establish $S'=\Server$ or even that the $X$ component is correct: the message
may have been tampered with.  But the session key secrecy theorem can be
applied.  Checking his nonce assures $B$ that~$K$ is a good key for talking
to~$A$, subject to the conditions of the secrecy theorem.

$B$'s guarantee follows from a lemma proved by induction.  It
resembles a regularity lemma.  Its main premise is that $B$'s certificate has
appeared, 
\[ \Crypt (\shrK B) \comp{\Nb, \Key K} \in\parts(\seespy), \]
with other premises and conclusion as in the guarantee itself.  Its proof is
complex, requiring several subsidiary lemmas:
\begin{itemize}
\item If the encrypted part of message~2 appears, then a suitable version of
  message~2 was actually sent.

\item The nonce~$\Nb$ uniquely identifies the other components of message~2's
  encrypted part.  This was discussed above~(\S\ref{sec:unique}).

\item A nonce cannot be used both as $\Na$ and as $\Nb$ in two
  protocol runs.  If $A\not\in\bad$ then the elements
\begin{align*}
  \Crypt (\shrK A) &\comp{\Na, \Agent A, \Agent B} \\
  \Crypt (\shrK A) &\comp{\Na', \Na, \Agent A', \Agent A}
\end{align*}
cannot both be in $\parts(\seespy)$.
\end{itemize}
The proof complexity arises from the use of nonces for binding and because
the two encrypted messages in step~3 have identical formats.

Now consider what $A$ (if uncompromised) can safely conclude upon receiving
message~4.  If a trace contains a message of the form
\[ \Says{B'}{A}{\comp{\Na, \Crypt (\shrK A) \comp{\Na, \Key K}}} \]
and if $A$ recalls sending message~1, 
\begin{align*}
   \Says{A}{B}{\comp{&\Na, \Agent A, \Agent B, \\
                     &\Crypt (\shrK A) \comp{\Na, \Agent A, \Agent B}}}
\end{align*}
then the server has sent a message of the correct form, for some~$\Nb$.  There
are many similarities, in both statement and proof, with $B$'s guarantee.  A
message, purportedly from~$B$, is considered as $A$ would see it.  Nonces are
compared with those from another message sent from~$A$ to~$B$.  The proof
again requires our proving by induction a lemma whose main premise is
\[ \Crypt (\shrK A) \comp{\Na, \Key K} \in\parts(\seespy), \]
with a detailed consideration of how nonces can be used.

\subsection{Proving a Simplified Protocol}\label{sec:simplified}

Abadi and Needham~\cite{abadi-prudent} suggest simplifying Otway-Rees by
eliminating the encryption in the first two messages.  Nonces serve only for
freshness, not for binding.  Message~3 explicitly names the intended
recipients.  
\begin{alignat*}{2}
  &1.&\; A\to B &: A, B, \Na \\
  &2.&\; B\to S &: A, B, \Na, \Nb \\
  &3.&\; S\to B &: \Na, \comp{\Na,A,B,\Kab}_{\Ka}, \comp{\Nb,A,B,\Kab}_{\Kb} \\
  &4.&\; B\to A &: \Na, \comp{\Na,A,B,\Kab}_{\Ka}
\end{alignat*}
The authors claim~\cite[page 11]{abadi-prudent},
`The protocol is not only more efficient but conceptually simpler after this
  modification.'
The machine proofs support their claims.  The vital guarantees to $B$ and~$A$,
from the last two messages, become almost trivial to prove.  Nonces do not
need to be unique and no facts need to be proved about them.  The new proof
script is smaller and runs faster.%
\footnote{From 82 to 40 seconds, and from 88 proof commands to 53.}

The new protocol is slightly weaker than the original.  The lack of encryption
in message~2 allows an intruder to masquerade as~$B$, though without learning
the session key.  The original Otway-Rees protocol assures~$A$ that $B$ is
present (I have proved this using Isabelle), but the new protocol does not.
However, the original version never assured $B$ that $A$ was present; anybody
could replay message~1, as Burrows et al.\ have noted \cite[page 247]{ban89}.

\section{A Public-Key Protocol: Needham-Schroeder}\label{sec:ns}

Needham-Schroeder is the obvious choice for demonstrating the inductive method
on public-key protocols.  Many researchers have investigated it, and Lowe has
discovered a subtle flaw~\cite{lowe-fdr}.  

\subsection{The Protocol and Lowe's Attack}\label{sec:protocol}

The full Needham-Schroeder protocol consists of seven steps, four of which are
devoted to distributing public keys.  Burrows et al.~\cite{ban89} identified a
flaw in this part of the protocol: there was no guarantee that the public keys
were fresh.  Assuming public keys to be universally known reduces the protocol
to three steps:
\begin{alignat*}{2}
  &1.&\quad  A\to B  &: \comp{\Na,A}_{\Kb} \\
  &2.&\quad  B\to A  &: \comp{\Na,\Nb}_{\Ka} \\
  &3.&\quad  A\to B  &: \comp{\Nb}_{\Kb}
\end{alignat*}
Message~2 assures $A$ of $B$'s presence, since only $B$ could have decrypted
$\comp{\Na,A}_{\Kb}$ to extract the freshly-invented nonce~$\Na$.  Similarly,
message~3 assures $B$ of $A$'s presence.  Burrows et al.\ claimed a
further property, namely that $\Na$ and~$\Nb$ become known only to
$A$ and~$B$.  (Such shared secrets might be used to compute a
session key.)  Lowe refuted this claim, noting that if $A$ ran the protocol
with an enemy~$C$, then $C$ could start a new run with any
agent~$B$, masquerading as $A$~\cite{lowe-fdr}.

One might argue that this is no attack at all.  An agent who is careless
enough to talk to the enemy cannot expect any guarantees.  The mechanized
analysis presented below reveals that the protocol's guarantees for~$A$ are
adequate.  However, those for $B$ are not: they rely upon $A$'s being careful,
which is a stronger assumption than mere honesty.  Moreover, the attack can
also occur if $A$ talks to an honest agent whose private key has been
compromised.  Lowe suggests a simple fix that provides good guarantees for
both $A$ and~$B$.

\subsection{Modelling the Protocol}

In the public-key model, an agent~$A$ has a public key~$\pubK A$, known to all
agents, and a private key~$\priK A$.  The spy knows the bad agents' private
keys.  No private key coincides with any public key.  In other respects, the
model resembles the shared-key one described above (\S\ref{sec:events}).

\begin{figure}[htbp]
\begin{ttbox}
\RuleLabel{Nil} [] \(\in\) ns_public

\RuleLabel{Fake} [| evs \(\in\) ns_public;  B\(\not=\)Spy;  
        X \(\in\) synth (analz (spies evs)) |]
     \(\Imp\) Says Spy B X  # evs \(\in\) ns_public

\RuleLabel{NS1} [| evs1 \(\in\) ns_public;  A\(\not=\)B;  Nonce NA \(\not\in\) used evs1 |]
     \(\Imp\) Says A B (Crypt (pubK B) \{|Nonce NA, Agent A|\})
           # evs1  \(\in\)  ns_public

\RuleLabel{NS2} [| evs2 \(\in\) ns_public;  A\(\not=\)B;  Nonce NB \(\not\in\) used evs2;
        Says A' B (Crypt (pubK B) \{|Nonce NA, Agent A|\}) \(\in\) set evs2 |]
     \(\Imp\) Says B A (Crypt (pubK A) \{|Nonce NA, Nonce NB|\})
           # evs2  \(\in\)  ns_public

\RuleLabel{NS3} [| evs3 \(\in\) ns_public;
        Says A  B (Crypt (pubK B) \{|Nonce{\ts}NA, Agent{\ts}A|\}) \(\in\) set evs3;
        Says B' A (Crypt (pubK A) \{|Nonce{\ts}NA, Nonce{\ts}NB|\}) \(\in\) set evs3 |]
     \(\Imp\) Says A B (Crypt (pubK B) (Nonce NB))
           # evs3  \(\in\)  ns_public
\end{ttbox}
\caption{Specifying the Needham-Schroeder Protocol} \label{fig:ns_public}
\end{figure}

Let us start with the original, flawed, Needham-Schroeder.
Figure~\ref{fig:ns_public} presents the inductive definition.  There are five
rules: three for the protocol steps and two standard ones, identical to those
in Fig.\ts\ref{fig:otway}.  There is no Oops message because the
protocol does not distribute session keys.  However, one could ask---as has
Meadows~\cite{meadows-ns}---what might happen if one of the nonces is
compromised.

More precisely, the protocol steps are as follows:
\begin{enumerate}
\item If, in the current trace, $\Na$ is a fresh nonce and $B$ is an agent
  distinct from~$A$, then we may add the event
\[ \Says{A}{B}{(\Crypt (\pubK B) \comp{\Na,A})}. \]

\item If the current trace contains an event of the form
\[ \Says{A'}{B}{(\Crypt (\pubK B) \comp{\Na,A})}, \]
and $\Nb$ is a fresh nonce and
$A\not=B$, then we may add the event 
\[ \Says{B}{A}{(\Crypt (\pubK A) \comp{\Na,\Nb})}. \]
Writing the sender as $A'$ means that $B$ does not 
know who sent the message.

\item If the current trace contains the two events
  \begin{align*}
    & \Says{A}{B}{\,(\Crypt (\pubK B) \comp{\Na,A})} \\
    & \Says{B'}{A}{(\Crypt (\pubK A) \comp{\Na,\Nb})}
  \end{align*}
then we may add the event 
\[ \Says{A}{B}{(\Crypt (\pubK B) \comp{\Nb})}. \] 
$A$ decrypts the message and checks that $\Na$ agrees with the nonce she
previously sent to~$B$.  She replies to $B$'s challenge by sending back~$\Nb$.
\end{enumerate}

As mentioned in \S\ref{sec:modelling}, we could model the implicit fourth
step, in which $B$ inspects the message arriving from~$A$.  But it suffices to
prove theorems stating what $B$ can infer from such an inspection, such as
that $A$ is present.

\subsection{Proving Guarantees for $A$}\label{sec:proving-A}

The guarantees for $A$ are that her nonce remains secret---from the spy---and
that $B$ is present.  The latter follows from the former, for if the spy does
not know $\Na$ then he could not have sent message~2.  The proofs require, as
lemmas, unicity properties for~$\Na$ saying that $\Na$ is only
used once.
\begin{itemize}
\item No value is ever used both as $\Na$ and as $\Nb$, even in separate
runs.
\item In any message of the form
$\Crypt (\pubK B) \comp{\Na,A}$, the value of nonce $\Na$ 
uniquely determines the agents $A$ and $B$, over all traffic.  
\end{itemize}
Both lemmas assume $\Na$ to be secret and form part of an inductive proof that
$\Na$ really is secret.  They hold because honest agents are specified to
choose unpredictable nonces with a negligible probability of collision.

The guarantee for~$A$ after step~2 is that the message indeed originated
with~$B$, provided it contains the expected nonce.  The guarantee is
consistent with Lowe's attack because, as always, it considers runs between
two uncompromised principals.  If $A$ runs the protocol with the spy then her
guarantee is void.  Lowe himself found no problem with the protocol from $A$'s
viewpoint~\cite[\S3.2]{lowe-fdr}; his attack concerns the guarantee
for~$B$.

\subsection{Proving Guarantees for $B$}\label{sec:proving-B}

The situation as seen by~$B$ is almost symmetrical to that seen by~$A$. 
Proving by induction that $\Nb$ remains secret would authenticate~$A$.  Most of
the Isabelle proof scripts for $A$'s theorems also work for~$B$ with
trivial alterations.  It is easy to prove that, if $\Nb$ is secret, then its
value in any message of the form $\Crypt (\pubK A) \comp{\Na,\Nb}$
uniquely determines $A$ and~$\Na$.

Unfortunately, $\Nb$ does not remain secret.  The attempt to prove its secrecy
fails, leaving a subgoal that contains (as a past event) $A$'s sending
message~1 to a compromised agent.  The subgoal describes a
consistent set of circumstances: Lowe's attack.  Details appear
in~\S\ref{sec:machine} below. 

Weaker properties can be proved.  If $A$ never sends $\Nb$ to anybody in step~3
of the protocol, then $\Nb$ remains secret.  In consequence, if $B$ receives
$\Nb$ in step~3 then $A$ has sent it, and is therefore present.  However, $A$
may have sent it to anybody.

The proof follows the usual argument (based on $A$'s
proofs), but assumes that $A$ says no messages of the form $\Crypt(\pubK C)\Nb$
for any~$C$.  With this additional assumption, $\Nb$ does remain secret; it
then follows that if $B$ sends $\Crypt (\pubK A) \comp{\Na,\Nb}$ as step~2 and
receives $\Crypt(\pubK B)\Nb$, then this reply came from~$A$.  Since this
conclusion contradicts the assumption, $B$ cannot receive $\Crypt(\pubK B)\Nb$.

The result above has the form $\neg Q$ implies $\neg P$, which is equivalent
to $P$ implies~$Q$.  If $B$ does receive $\Crypt(\pubK B)\Nb$, then
$A$ has indeed sent the message $\Crypt(\pubK C)\Nb$ for some~$C$.

This example suggests a general strategy to prove that decrypting a
message of the form
$\Crypt (\pubK A) X$ indicates $A$'s presence.  Prove that if $A$ never
performs the step in which that message is decrypted, then some item in~$X$
remains secret.  Conclude that if $X$ is revealed then $A$ must have performed
the decryption.

This roundabout procedure is necessary because the mere act of decryption
gives weaker guarantees than exhibiting a signed message.  Consider the
following protocol:
\begin{alignat*}{2}
  &1.&\quad  A\to B  &: \Na,A \\
  &2.&\quad  B\to A  &: \comp{\Na}_{\Kb^{-1}}, \Nb \\
  &3.&\quad  A\to B  &: \comp{\Nb}_{\Ka^{-1}}
\end{alignat*}
The nonces are broadcast to the world, but the signatures obviously assure $A$
and~$B$ of the other's presence.

\subsection{A Glimpse at the Machine Proofs}\label{sec:machine}

To give an impression of the Isabelle formalization, Fig.\ts\ref{fig:ns-thms}
presents the theorems providing guarantees for~$A$.  They are numbered
as follows.

\begin{figure} 
\begin{ttbox}
\textrm{1}  [| Crypt (pubK B) \{|Nonce NA, Agent A|\}\(\,\in\,\)parts(spies evs);
      Nonce NA\(\,\not\in\,\)analz(spies evs);
      evs\(\,\in\,\)ns_public |]
   \(\,\Imp\,\) Crypt (pubK C) \{|NA', Nonce NA|\}\(\,\not\in\,\)parts(spies evs)

\textrm{2}  [| Nonce NA\(\,\not\in\,\)analz(spies evs);  evs\(\,\in\,\)ns_public |]
   \(\,\Imp\,\) \(\exists\,\)A' B'. \(\forall\,\)A B.
          Crypt (pubK B) \{|Nonce NA, Agent A|\}\(\,\in\,\)parts(spies evs)
            \(\,\imp\,\) A=A' & B=B'

\textrm{3}  [| Crypt(pubK B)  \{|Nonce NA, Agent A|\} \(\,\in\,\)parts(spies evs); 
      Crypt(pubK B') \{|Nonce NA, Agent A'|\}\(\,\in\,\)parts(spies evs); 
      Nonce NA\(\,\not\in\,\)analz(spies evs);                            
      evs\(\,\in\,\)ns_public |] 
  \(\,\Imp\,\) A=A' & B=B'

\textrm{4}  [| Says A B (Crypt (pubK B) \{|Nonce NA, Agent A|\})\(\,\in\,\)set evs;
      A\(\,\not\in\,\)bad;  B\(\,\not\in\,\)bad;  evs\(\,\in\,\)ns_public |]   
   \(\,\Imp\,\) Nonce NA\(\,\not\in\,\)analz(spies evs)

\textrm{5}  [| Says A  B (Crypt (pubK B) \{|Nonce NA, Agent A|\}) \(\,\in\,\)set evs;
      Says B' A (Crypt (pubK A) \{|Nonce NA, Nonce NB|\})\(\,\in\,\)set evs;
      A\(\,\not\in\,\)bad;  B\(\,\not\in\,\)bad;  evs\(\,\in\,\)ns_public |]
  \(\,\Imp\,\) Says B A (Crypt(pubK A) \{|Nonce NA, Nonce NB|\})\(\,\in\,\)set evs
\end{ttbox}
\caption{The Guarantees for $A$ in Isabelle/HOL Notation}
 \label{fig:ns-thms}
\end{figure}

\begin{enumerate}
\item This unicity lemma states that $\Na$ (if secret) is not also used as
$\Nb$.  It is proved by induction. 

\item This unicity lemma states that, if $\Na$ is secret, then its appearance
in any instance of message~1 determines the other components.  It too follows
by induction, with a standard proof script.

\item This corollary of the previous lemma has a trivial proof.

These unicity lemmas refer to the presence of encrypted messages
anywhere in past traffic.  The remaining theorems refer to events of the form 
$\Says{A}{B}{X}$ involving such encrypted messages.

\item This crucial theorem guarantees the secrecy of~$\Na$.  The conditions
  $A\not\in\bad$ and $B\not\in\bad$ express that both $A$ and $B$ are
  uncompromised.  The proof is by induction; it relies on the previous three
  lemmas, which assume the secrecy of~$\Na$ as an induction hypothesis.
  
\item This theorem is $A$'s final guarantee.  If $A$ has used $\Na$ to start a
  run with~$B$ and receives the message $\Crypt (\pubK A) \comp{\Na,\Nb}$, then
  $B$ has sent that message.  It is subject to both agents' being
  uncompromised.  The proof is by induction and relies on the secrecy and
  unicity of $\Na$.
\end{enumerate}
The proof script for all five theorems comprises 27 commands (tactic
invocations) and executes in ten seconds, or two seconds per theorem.

What about the guarantees for~$B$?  Attempting to prove the secrecy of~$\Nb$
leads to a subgoal that appears to have no proof.
\begin{ttbox}
[| A\(\,\not\in\,\)bad;  B\(\,\not\in\,\)bad;  C\(\,\in\,\)bad;  
   evs3\(\,\in\,\)ns_public;  
   Says A  C (Crypt (pubK C) \{|Nonce NA, Agent A|\})  \(\,\in\,\) set evs3;
   Says B' A (Crypt (pubK A) \{|Nonce NA, Nonce NB|\}) \(\,\in\,\) set evs3;
   Says B  A (Crypt (pubK A) \{|Nonce NA, Nonce NB|\}) \(\,\in\,\) set evs3;
   Nonce NB\(\,\not\in\,\)analz (spies evs3) |]
==> False
\end{ttbox}
This situation might arise when the last event is an instance of step~3, as we
can tell because the trace is called $evs3$.  Agents $A$ and~$B$ are
uncompromised and~$A$ has used $\Na$ to start a run with a compromised
agent,~$C$.  Somebody has sent the message $\Crypt (\pubK A) \comp{\Na,\Nb}$.  We
must show that these circumstances are contradictory, since the conclusion is
just \textsf{False}. The conclusion is the simplified form of the claim that
$\Nb$ remains secret even after $A$ has sent the step~3 message $\Crypt (\pubK
C) \comp{\Nb}$, but this message reveals~$\Nb$ to the spy.

Such proof states can be hard to interpret.  Does the induction formula require
strengthening?  Must additional lemmas be proved?  But, in this case, we easily
recognize Lowe's attack.  The assumptions describe events that could actually
occur: $\Nb$ need not remain secret.

\subsection{Analyzing the Strengthened Protocol}\label{sec:ns-public}

Lowe~\cite{lowe-fdr} suggests improving the Needham-Schroeder protocol by
adding explicitness.  In step~2, agent $B$ includes his identity:
\begin{alignat*}{2}
  &1.&\quad  A\to B  &: \comp{\Na,A}_{\Kb} \\
  &2.&\quad  B\to A  &: \comp{\Na,\Nb,B}_{\Ka} \\
  &3.&\quad  A\to B  &: \comp{\Nb}_{\Kb}
\end{alignat*}
The previous proof scripts, by and large, still work for this version.  Thanks
to Isabelle's high level of automation, minor changes such as that above
seldom interfere with existing proofs.  The guarantees for~$A$ are proved
precisely as before.

In proving guarantees for~$B$, we naturally seek to strengthen them.  The
unicity property for~$\Nb$ states that, if $\Nb$ is secret, then its presence in
step~2 uniquely determines all other message components (recall
\S\ref{sec:proving-B}).  Step~2 now has the form
\[ \Crypt (\pubK A) \comp{\Na,\Nb,B}. \]
Nonce~$\Nb$ determines not only $A$ and~$\Na$ but also~$B$.  This additional
fact lets us prove the secrecy of $\Nb$.  Recall the subgoal presented
in~\S\ref{sec:machine}.  With the new version of the protocol, somebody has
sent the message
\[ \Crypt (\pubK A) \comp{\Na,\Nb,C}. \]
Also, agent $B$ has sent the message
\[ \Crypt (\pubK A) \comp{\Na,\Nb,B}. \]
The unicity theorem for~$\Nb$ implies $B=C$, a contradiction because $C$ is
compromised and $B$ is not.

\section{A Recursive Protocol}\label{sec:recur}

This protocol~\cite{bull-otway} generalizes Otway-Rees to an arbitrary number
of parties.  First, $A$ contacts~$B$.  If $B$ then contacts the authentication
server then the run resembles Otway-Rees.  But $B$ may choose to contact some
other agent~$C$, and so forth; a chain of arbitrary length may form.  During
each such round, an agent adds its name and a fresh nonce to an ever-growing
request message.

For the sake of discussion, suppose that $C$ does not extend the chain but
instead contacts the authentication server.  The server generates fresh
session keys $\Kab$ and $\Kbc$---in the general case, one key for each pair of
agents adjacent in the chain.  It encloses each session key in two
certificates, one for each party, and gives the bundle to~$C$.  Each agent
removes two certificates and forwards the rest to its predecessor in the
chain.  Finally, $A$ receives one certificate, containing~$\Kab$.

Such a protocol is hard to specify, let alone analyze.  The number of steps,
the number of parties and the number of session keys can vary.  The server's
response to the agents' accumulated requests must be given as a recursive
program.

Even which properties to prove are not obvious.  One might simplify the
protocol to distribute a single session key, common to all the agents in the
chain.  But then, security between $A$ and~$B$ would depend upon the honesty
of~$C$, an agent possibly not known to~$A$.  There may be applications where
such a weak guarantee might be acceptable, but it seems better to give a
separate session key to each adjacent pair.  

I have proved a general guarantee for each participant.  If it receives a
certificate containing a session key and the name of another agent, then the
spy will never know the key.  The Isabelle proofs are modest in scale.  Fewer
than 30 results are proved, using under 130 commands; they run in about
three minutes.

\subsection{The Recursive Authentication Protocol}\label{sec:recur-auth}

\begin{figure*}
\begin{center}
\epsfig{file=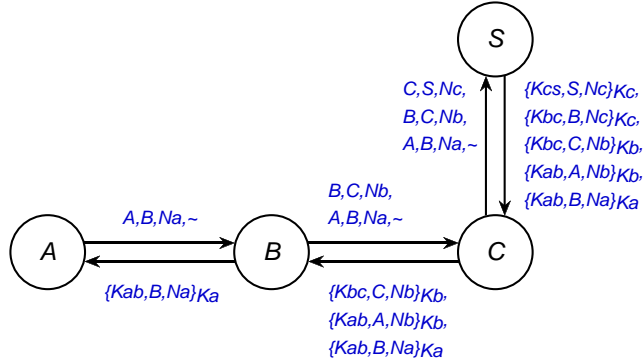,width=20pc}
\end{center}
\caption{The Recursive Authentication Protocol with Three Clients} \label{fig:recur-eps}
\end{figure*}

The protocol was invented by John Bull of APM Ltd.   In the description below,
let
$\Hash X$ be the hash of~$X$ and $\Hash_X Y$ the pair
$\comp{\Hash\comp{X,Y},Y}$.  Typically, $X$ is an agent's long-term shared key
and $\Hash\comp{X,Y}$ is a message digest, enabling the server to check that
$Y$ originated with that agent.  Figure~\ref{fig:recur-eps} shows a typical
run, omitting the hashing.

Agent~$A$ starts a run by sending $B$ a request:
\[  1.\;  A\to B  : \Hash_{\Ka} \comp{A,B,\Na,{-}} \]
Here $\Ka$ is $A$'s long-term shared key, $\Na$ is a fresh nonce, and $(-)$ is
a placeholder indicating that this message started the run.  In response, $B$
sends something similar but with $A$'s message in the last position:
\[  2.\;  B\to C  : \Hash_{\Kb} \comp{B,C,\Nb,
                                 \overbrace{\Hash_{\Ka}\comp{A,B,\Na,{-}}}
                                    ^{\text{from }A}} 
\]
Step~2 may be repeated as many times as desired.  Each time, new components
are added to the message and a new message digest is prefixed.  The recursion
terminates when some agent performs step~2 with the server as the
destination.

In step~3, the server prepares session keys for each caller-callee pair.  It
traverses the accumulated requests to build up its response.  If (as
in~\S\ref{sec:recur-auth}) the callers were $A$, $B$ and~$C$ in that order,
then the final request is 
\newcommand\point[1]{\begin{array}[t]{@{}c@{}}#1\\ \uparrow\end{array}}
\[ \Hash_{\Kc} \comp{\point{C},S,\Nc, \Hash_{\Kb} \comp{B,\point{C},\Nb,
      \Hash_{\Ka}\comp{A,B,\Na,{-}}}}. \]
The arrows point to the occurrences of~$C$, which appear in the
outer two levels.  $C$ has called~$S$ (the server)
and was called by~$B$.  The server generates session keys $\Kcs$ and $\Kbc$ and
prepares the certificates $\comp{\Kcs,S,\Nc}_{\Kc}$ and $\comp{\Kbc,B,\Nc}_{\Kc}$.
The session key $\Kcs$ is redundant because $C$ already shares $\Kc$ with the
server.  Including it allows the last agent in the chain to be treated like all
other agents except the first: the initiator receives only one session key.

Having dealt with $C$'s request, the server discards it.  Looking at the
remaining outer two levels, the request message is
\[ \Hash_{\Kb} \comp{\point{B},C,\Nb,\Hash_{\Ka}\comp{A,\point{B},\Na,{-}}}. \]
The server now prepares two certificates for~$B$, namely $\comp{\Kbc,C,\Nb}_{\Kb}$
and $\comp{\Kab,A,\Nb}_{\Kb}$.  Note that $\Kbc$ appears in two certificates, one
intended for~$C$ (containing nonce~$\Nc$ and encrypted with key~$\Kc$) and one
for~$B$.

At the last iteration, the request message contains only one level:
\[ \Hash_{\Ka}\comp{\point{A},B,\Na,{-}}. \]
The $(-)$ token indicates the end of the requests.  The server generates one
session key and certificate for~$A$, namely $\comp{\Kab,B,\Na}_{\Ka}$.

Having processed the request message, the server returning a bundle of
certificates.  In our example, it would return five certificates to~$C$.
\begin{alignat*}{2}
  &3.&\;  S\to C :\, & \comp{\Kcs,S,\Nc}_{\Kc}, \comp{\Kbc,B,\Nc}_{\Kc},\\
                   &&& \comp{\Kbc,C,\Nb}_{\Kb}, \comp{\Kab,A,\Nb}_{\Kb},\\
                   &&& \comp{\Kab,B,\Na}_{\Ka} 
\end{alignat*}

In step~4, an agent accepts the first two certificates and forwards the rest to
its predecessor in the chain.  Every agent performs this step except the
one who started the run.
\begin{alignat*}{2}
  &4.&\;  C\to B  :\,& \comp{\Kbc,C,\Nb}_{\Kb}, \comp{\Kab,A,\Nb}_{\Kb}, \\
      &&& \comp{\Kab,B,\Na}_{\Ka}\\
  &4'.&\;  B\to A  :\,& \comp{\Kab,B,\Na}_{\Ka}
\end{alignat*}

The description above describes a special case: a protocol run with three
clients.  The conventional protocol notation cannot cope with arbitrary
numbers of participants, let alone recursive processing of nested messages.
Section~\S\ref{sec:model} below will specify the protocol as an inductive
definition.

\subsection{Deviations from the Protocol}

I have corrected a flaw in the original protocol.  My formalization of the
protocol differs from the original in some other respects, as well.

In the original protocol, an agent's two certificates are distinguished only
by their order of arrival; an intruder could easily exchange them.  To correct
this flaw, I have added the other party's name to each certificate.  Such
explicitness is good engineering practice~\cite{abadi-prudent}.  It also
simplifies the proofs (recall \S\ref{sec:simplified}).  Bull and Otway have
accepted my change to their protocol~\cite{bull-otway}.
  
The dummy session key~$\Kcs$ avoids having to treat the last agent as a special
case.  All agents except the first take two certificates.  An implementation
can safely omit the dummy certificate.  Removing information from the system
makes less information available to an intruder.
  
The original protocol implements encryption using exclusive-or (XOR) and
hashing.  For verification purposes, encryption should be taken as primitive.
Correctness of the protocol does not depend upon the precise form of
encryption, provided it is implemented properly; the original use of
XOR was flawed (see \S\ref{sec:attacks}).

Protocol certificates are accompanied by agent names sent in clear.  It is
safe to simplify the specification by omitting these names.

\subsection{Modelling the Protocol}\label{sec:model}

Requests in the protocol have the form $\Hash_X Y$, where $Y$ may contain
another request.  The $\Hash_X Y$ notation for message digests is trivially
defined in Isabelle:
\[ \Hash[X] Y\equiv \comp{\Hash\comp{X,Y},Y}. \]
Most proofs do not apply this definition directly.  The default rewrite rules
apply only if $\Hash_X Y$ appears, say, in the argument of $\parts$, where the
expression can be simplified.  Such rules help prevent exponential blowup.

A further law is subject to $X\not\in\synth(\analz H)$, as when
$X$ is an uncompromised long-term key:
\begin{multline*}
 \Hash_X Y\in\synth(\analz H) \iff \\
                      \Hash\comp{X,Y}\in\analz H\conj Y\in\synth(\analz H)
\end{multline*}
The message $\Hash_X Y$ can be spoofed iff $Y$ can be and a
suitable message digest is available (an unlikely circumstance).  

\begin{figure*}[htbp]
\begin{alltt*}\footnotesize
\RuleLabel{Nil} [] \(\in\) recur

\RuleLabel{Fake} [| evs \(\in\) recur;  B\(\not=\)Spy;  
        X \(\in\) synth (analz (spies evs)) |]
     \(\Imp\) Says Spy B X  # evs \(\in\) recur

\RuleLabel{RA1} [| evs1 \(\in\) recur;  A\(\not=\)B;  A\(\not=\)Server;  Nonce NA \(\not\in\) used evs1 |]
     \(\Imp\) Says A B 
           (Hash[Key(shrK A)] 
            \{|Agent A, Agent B, Nonce NA, Agent Server|\})
         # evs1 \(\in\) recur

\RuleLabel{RA2} [| evs2 \(\in\) recur;  B\(\not=\)C;  B\(\not=\)Server;  Nonce NB \(\not\in\) used evs2;
        Says A' B PA \(\in\) set evs2 |]
     \(\Imp\) Says B C (Hash[Key(shrK B)]\{|Agent B, Agent C, Nonce NB, PA|\})
         # evs2 \(\in\) recur

\RuleLabel{RA3} [| evs3 \(\in\) recur;  B\(\not=\)Server;
        Says B' Server PB \(\in\) set evs3;
        (PB,RB,K) \(\in\) respond evs3 |]
     \(\Imp\) Says Server B RB # evs3 \(\in\) recur

\RuleLabel{RA4} [| evs4 \(\in\) recur;  A\(\not=\)B;  
        Says B  C \{|XH, Agent B, Agent C, Nonce NB, 
                    XA, Agent A, Agent B, Nonce NA, P|\} 
          \(\in\) set evs4;
        Says C' B \{|Crypt (shrK B) \{|Key KBC, Agent C, Nonce NB|\},
                    Crypt (shrK B) \{|Key KAB, Agent A, Nonce NB|\}, 
                    RA|\}
          \(\in\) set evs4 |]
     \(\Imp\) Says B A RA # evs4 \(\in\) recur
\end{alltt*}
\caption{Specifying the Recursive Protocol} \label{fig:recur}
\end{figure*}

For the most part, the protocol is modelled just like the fixed-length
protocols discussed above.  Figure~\ref{fig:recur} presents the inductive
definition.  The rules for the empty trace and the spy are standard.  The
other rules can be paraphrased as follows:
\begin{enumerate}
\item If, in the current trace, $\Na$ is a fresh nonce and $B$ is an agent
  distinct from~$A$ and~$\Server$, then we may add the event
\[ \Says{A}{B}{(\Hash_{\shrK A}\comp{A,B,\Na,{-}})}. \]
$A$'s long-term key is written~$\shrK A$.
For the token $(-)$ I used the name $\Server$, but any fixed message would do
as well.

\item If the current trace contains the event $\Says{A'}{B}{Pa}$, where
  $Pa=\comp{Xa,A,B,\Na,P}$, and $\Nb$ is a fresh nonce and $B\not=C$, then we
  may add the event
\[ \Says{B}{C}{(\Hash_{\shrK B}\comp{B,C,\Nb,Pa})}. \]
The variable $Xa$ is how $B$ sees $A$'s hash value; he does not have the
key needed to verify it.  Component~$P$ is $(-)$ if $A$ started the
run or might have the same form as~$Pa$, nested to any depth.  Agent $C$ might
be the server or anybody else.

The specification actually omits the equation defining~$Pa$.  It appears to be
unnecessary, and its omission simplifies the proofs.  They therefore hold of a
weaker protocol in which any agent may react to any message by sending an
instance of step~2.  Ill-formed requests may result, but the server will
ignore them.

\item If the current trace contains the event 
  $\Says{B'}{\Server}{Pb}$, and $B\not=\Server$, and if the server can build
  from request~$Pb$ a response~$Rb$, then we may add the event
\[ \Says{\Server}{B}{Rb}. \]
The construction of $Rb$ includes verifying the integrity of~$Pb$; this
process is itself defined inductively, as we shall see.  The rule does not
constrain the agent~$B$, allowing the server to send the response to anybody. 
We could get the right value of~$B$ from~$Pb$, but the proofs do not require
such details.

\item If the current trace contains the two events
\begin{align*}
    \SAYS\,{B}\;{C}  &\,{(\Hash_{\shrK B}\comp{B,C,\Nb,Pa})} \\
    \SAYS\,{C'}\,{B} &\,{\comp{\begin{array}[t]{@{}l@{}}
                        \Crypt(\shrK B)\comp{\Kbc,C,\Nb},\,\\
                        \Crypt(\shrK B)\comp{\Kab,A,\Nb},\,  R}
                        \end{array}}
\end{align*}
and $A\not=B$, then we may add the event 
\[ \Says{B}{A}{R}. \]
$B$ decrypts the two certificates, compares their nonces with the value
of $\Nb$ he used, and forwards the remaining certificates ($R$).
\end{enumerate}

The final step of the protocol is the initiator's acceptance of the last
certificate, $\Crypt(\shrK A)\comp{\Kab,B,\Na}$.  This implicit step need not
be modelled; all certificates will be proved to be authentic.

There is no Oops message (recall \S\ref{sec:oops}).  It cannot easily be
expressed for the recursive authentication protocol because a key never
appears together with both its nonces.  The spy can still get hold of session
keys using the long-term keys of compromised agents.

\subsection{Modelling the Server}

The server creates the list of certificates according to another inductive
definition.  It defines not a set of traces but a set of triples $(P,R,K)$
where $P$ is a request, $R$ is a response and $K$ is a session key.  Such
triples belong to the set $\respond evs$, where $evs$ (the current trace) is
supplied to prevent the reuse of old session keys.  Component~$K$ returns the
newest session key to the caller for inclusion in a second certificate.

\begin{figure*}[htbp]
\begin{alltt*}\footnotesize
\RuleLabel{One} [| A \(\not=\) Server;  Key KAB \(\not\in\) used evs |]
     \(\Imp\) (Hash[Key(shrK A)] \{|Agent A, Agent B, Nonce NA, Agent Server|\}, 
          \{|Crypt(shrK A) \{|Key KAB, Agent B, Nonce NA|\}, Agent Server|\},
          KAB)   \(\in\) respond evs

\RuleLabel{Cons} [| (PA, RA, KAB) \(\in\) respond evs;  
        Key KBC \(\not\in\) used evs;  Key KBC \(\not\in\) parts \{RA\};
        PA \(=\) Hash[Key(shrK A)] \{|Agent A, Agent B, Nonce NA, P|\};
        B \(\not=\) Server |]
     \(\Imp\) (Hash[Key(shrK B)] \{|Agent B, Agent C, Nonce NB, PA|\}, 
          \{|Crypt (shrK B) \{|Key KBC, Agent C, Nonce NB|\}, 
            Crypt (shrK B) \{|Key KAB, Agent A, Nonce NB|\},
            RA|\},
          KBC)   \(\in\) respond evs
\end{alltt*}
\caption{Specifying the Server} \label{fig:respond}
\end{figure*}

The occurrences of $\Hash$ in the definition ensure that the server accepts
requests only if he can verify the hashes using his knowledge of the long-term
keys.  The inductive definition (Fig.\ts\ref{fig:respond}) consists of two
cases.

\begin{enumerate}
\item If $\Kab$ is a fresh key (that is, not used in $evs$) then 
\begin{align*}
   (&\Hash_{\shrK A}\comp{A,B,\Na,{-}},\\
    &\comp{\Crypt(\shrK A)\comp{\Kab,B,\Na},\,{-}},\\
    &\Kab) \quad\in\quad\respond evs.
\end{align*}
This base case handles the end
of the request list, where $A$ seeks a session key with~$B$.

\item If $(Pa,Ra,\Kab)\in\respond evs$ and $\Kbc$ is fresh (not
used in $evs$ or~$Ra$) and 
\[ Pa=\Hash_{\shrK A}\comp{A,B,\Na,P} \]
then 
\begin{align*}
   (&\Hash_{\shrK B}\comp{B,C,\Nb,Pa},\\
    &\comp{\Crypt(\shrK B)\comp{\Kbc,C,\Nb},\,\\
    &\phantom{\lbb} \Crypt(\shrK B)\comp{\Kab,A,\Nb},\, Ra},\\
    &\Kbc) \quad\in\quad\respond evs.
\end{align*}
The recursive case handles a request list where $B$ seeks a session key
with~$C$ and has himself been contacted by~$A$.
The $\respond$ relation is best understood as a pure Prolog program. 
Argument $Pa$ of $(Pa,Ra,\Kab)$ is the input, while $Ra$ and $\Kab$ are outputs. 
Key~$\Kab$ has been included in the response~$Ra$ and must be included in one of
$B$'s certificates too.
\end{enumerate}

An inductive definition can serve as a logic program.  Because the concept
is Turing powerful, it can express the most complex behaviours.  Such
programs are easy to reason about.

\subsubsection*{A Coarser Model of the Server}\label{sec:coarser}

For some purposes, $\respond$ is needlessly complicated.  Its
input is a list of $n$ requests, for $n>0$, and its output is a list of $2n+1$
certificates.  Many routine lemmas hold for any list of certificates of the
form $\Crypt(\shrK B)\comp{K,A,N}$.  The
inductive relation $\responses$ generates the set of all such lists.  It
contains all possible server responses and many impossible ones.

The base case is simply $({-})\in\responses evs$
and the recursive case is
\[ \comp{\Crypt(\shrK B)\comp{K,A,N},\,R}\in\responses evs \]
if $R\in\responses evs$ and $K$ is not used in $evs$.

In secrecy theorems (those expressed in terms of $\analz$), each occurrence of
$\Crypt$ can cause a case split, resulting in a substantial blowup after
simplification.  Induction over $\responses$ introduces only one $\Crypt$, but
induction over $\respond$ introduces three.  Because $\responses$ includes
invalid outputs, some theorems can only be proved for
$\respond$.

\subsection{Main Results Proved}\label{sec:results}

For the most part, the analysis resembles that of the Otway-Rees protocol.
Possibility properties are proved first, then regularity lemmas.  Secrecy
theorems govern the use of session keys, leading to the session key secrecy
theorem: if the certificate $\Crypt(\shrK A)\comp{\Kab,B,\Na}$ appears as part
of any traffic, where $A$ and $B$ are uncompromised, then $\Kab$ will never
reach the spy.  Another theorem guarantees that such certificates
originate only with the server.

Possibility properties are logically trivial.  All they tell us is that the
rules' message formats are compatible.  However, their machine proofs require
significant effort (or computation) due to the complexity of the terms that
arise and the number of choices available.  I proved cases corresponding to
runs with up to three agents plus the server and spy.  General theorems for
$n$ agents could be proved by induction on~$n$, but the necessary effort
hardly seems justified.

A typical regularity lemma states that the long-term keys of uncompromised
agents never form part of any message.  They do form part of hashed messages,
however; recall the discussion in~\S\ref{sec:hashing} above.

Security properties are proved, as always, by induction over the protocol
definition.  For this protocol, the main inductive set ($\recur$) is defined in
terms of another ($\respond$).  All but the most trivial proofs require
induction over both definitions.

An easily-proved result lets us reduce $\responses$ to $\respond$, 
justifying the use of induction over $\responses$:
\[ \infer{Rb \in \responses evs}{(Pa,Rb,\Kab)\in \respond evs}  \]

Most results are no harder to prove than for a fixed-length protocol.  Proving
a theorem requires four commands on average, of which two are quite
predictable: induction and simplification.  The outer induction yields six
subgoals: one for each protocol step, plus the base and fake cases.  The inner
induction replaces the step~3 case by two subgoals: the server's base
case and inductive step.  Few of these seven subgoals survive simplification.
Only the theorems described below have difficult proofs.

Nonces generated in requests are unique.  There can be
at most one hashed value containing the key of an uncompromised agent
($A\not\in\bad$) and any specified nonce value,~$\Na$.
\begin{multline*}
\exists B'\, P'.\, \forall B\,P. \\
    \Hash\comp{\Key(\shrK A),\Agent A,\Agent B,\Na,P}
    \in\parts(\seespy) \\
\imp B=B' \conj P=P'.
\end{multline*}
Although it is not used in later proofs, this theorem is important.  It
lets agents identify runs by their nonces.  The theorem applies to all
requests, whether generated in step~1 or step~2.  For the Otway-Rees protocol,
each of the two steps requires its own theorem.  The reasoning here is
similar, but one theorem does the work of two, thanks to the protocol's
symmetry.  The nesting of requests does not affect the reasoning.

The session key compromise theorem is formulated just as for
Otway-Rees (see \S\ref{sec:proving-secrecy}), but its proof is much more
difficult.  The inner induction over $\respond$ leads to excessive case
splits.  It was to simplify this proof that I defined the set $\responses$.

Unicity for session keys is unusually complicated because each key
appears in two certificates.  Moreover, the certificates are created in
different iterations of $\respond$.  The unicity
theorem states that, for any~$K$, if there is a certificate of the form
\[ \Crypt(\shrK A)\comp{K,B,\Na} \]
(where $A$ and $B$ are uncompromised) then the
only other certificate containing~$K$ must have the form 
\[ \Crypt(\shrK B)\comp{K,A,\Nb}, \]
for some~$\Nb$.  If $(PB,RB,K)\in\respond evs$ then
\begin{multline*}
\exists A'\, B'.\, \forall A\,B\,N. \\
    \Crypt (\shrK A) \comp{\Key K, \Agent B, N}
    \in\parts\{RB\} \\
\imp (A'=A \conj B'=B) \disj (A'=B \conj B'=A).
\end{multline*}
This theorem seems quite strong.  An agent who receives a certificate
immediately learns which other agent can receive its mate, subject to the
security of both agents' long-term keys.  One might hope that the session key
secrecy theorem would follow without further ado.  The only messages
containing session keys contain them as part of such certificates, and thus
the keys are safe from the spy.  But such reasoning amounts to another
induction over all possible messages in the protocol.  The theorem must be
stated (stipulating $A$,~$A'\not\in\bad$) and proved:
\begin{multline*}
\Crypt (\shrK A) \comp{\Key K, \Agent A', N}
    \in\parts(\seespy) \\
\imp \Key K \not\in \analz(\seespy) 
\end{multline*}
The induction is largely straightforward except for the step~3 case.  The inner
induction over $\respond$ leads to such complications that it must be proved
beforehand as a lemma.  If $(PB,RB,\Kab)\in\respond evs$ then
\begin{multline*}
\forall A\, A'\, N.\, A \not\in\bad \conj A' \not\in\bad \\
     \imp \Crypt (\shrK A) \comp{\Key K, \Agent A', N} \in\parts\{RB\} \\
        \imp \Key K \not\in \analz (\INS{RB}\seespy)
\end{multline*}
Although each session key appears in two certificates, they both have the same
format.  A single set of proofs applies to all certificates.  Once again, the
protocol's symmetry halves the effort compared with Otway-Rees.

\goodbreak
It may be instructive to see some theorems in Isabelle syntax.  Here is the
session key compromise theorem:
\begin{ttbox}
[| evs \(\in\) recur;  KAB \(\not\in\) range shrK |] 
\(\Imp\) (Key K \(\in\) analz (insert (Key KAB) (spies evs))) \(=\)     
    (K \(=\) KAB \(\disj\) Key K \(\in\) analz (spies evs))
\end{ttbox}
And here is the session key secrecy theorem:
\begin{ttbox}
[| Crypt (shrK A) \{|Key K, Agent A', N|\} \(\in\) parts (spies evs);
   A \(\not\in\) bad;  A' \(\not\in\) bad;  evs \(\in\) recur |]  
\(\Imp\) Key K \(\not\in\) analz (spies evs)
\end{ttbox}

\subsection{Potential Attacks}\label{sec:attacks}

All proofs are subject to the assumptions implicit in the model.  Attacks
against the protocol or implementations of it can still be expected.  One
`attack' is obvious: in step~2, agent $B$ does not know whether $A$'s
message is recent; at the conclusion of the run, $B$ still has no evidence
that $A$ is present.  The spy can masquerade as $A$ by replaying an old
message of hers, but cannot read the resulting certificate without her
long-term key.

Allowing type confusion (such as passing a nonce as a key) often admits
attacks~\cite{lowe-casper,meadows-ns} in which one form of certificate is
mistaken for another.  The recursive authentication protocol is safe from
such attacks because it has only one form of certificate.  However, encryption
must be secure.

In the original protocol, each session key was encrypted by forming its XOR
with a hash value, used as a one-time pad.  Unfortunately, each hash value was
used twice: $B$'s session keys $\Kab$ and $\Kbc$ were encrypted as
\[ \Kab\oplus \Hash\comp{\Kb,\Nb} \quad\text{and}\quad
   \Kbc\oplus \Hash\comp{\Kb,\Nb}. 
\]
By forming their XOR, an eavesdropper could immediately obtain $\Kab\oplus
\Kbc$, $\Kbc\oplus \Kcd$, etc.  Compromise of any one session key would reveal
all the others~\cite{ryan-attack}.

\section{Related Work}\label{sec:related}

Several other researchers are using inductive or trace models.  Verification
is done using general-purpose theorem provers or model checkers, or by hand.

\newcommand\KeyAB{\mathop{\mathsf{KeyAB}}} 

In early work, Kemmerer~\cite{kemmerer-analyzing} analyzed a protocol in the
Ina Jo specification language, which is based on first-order logic.  Using an
animation tool, he identified two weaknesses in the protocol.  He modelled the
system as an automaton, defining the initial state and the state
transformations, and specifying security goals as invariants.  Proving that
state transformations preserve an invariant is the same style of reasoning as
induction.  Gray and McLean~\cite{gray-temporal} also establish a security
invariant by induction, though their work is based on temporal logic and their
proofs are done by hand.

Bolignano's work~\cite{boli-approach} is based on the Coq proof checker.  His
$X\mathbin{\mathsf{known\_in}}H$ is equivalent to my $X\in\synth(\analz H)$ and
models fraudulent messages. Instead of formalizing traces, he models the states
of the four agents
$\mathsf{A}$, $\mathsf{B}$, $\Server$ and the spy.  He and
M{\'e}nissier-Morain~\cite{boli-Coq} have formalized the Otway-Rees protocol.
In their model, the server uses the function $\KeyAB$ to choose session keys:
whenever $\mathsf{A}$ and $\mathsf{B}$ participate in a run, the server issues
$\KeyAB(\mathsf{A},\mathsf{B})$.  They have proved three properties.  The
first resembles the forwarding lemmas described in~\S\ref{sec:forwarding}.
The second states that the secret keys $\Ka$ and $\Kb$ remain secret;
it is similar to a regularity lemma described in~\S\ref{sec:regularity}.  The
third property states that $\KeyAB(\mathsf{A},\mathsf{B})$ cannot be decrypted
from traffic even with the help of all other session keys.

Their model is somewhat restrictive.  The constants $\mathsf{A}$ and
$\mathsf{B}$ are fixed in the roles of initiator and responder, respectively.
The spy starts off holding no keys, which gives him no prospect of
impersonating honest agents or decrypting messages.  Also, note
that the attack of
\S\ref{sec:attack} works not by giving $\KeyAB(\mathsf{A},\mathsf{B})$ to the spy, but by getting
$\mathsf{A}$ to accept $\KeyAB(\mathsf{C},\mathsf{A})$ as a good key for
talking to~$\mathsf{B}$.

Lowe's approach is based on traces, which are specified using the process
calculus CSP~\cite{hoare85} and examined using the model-checker FDR\@.  His
work originates in that of Roscoe~\cite{roscoe-modelling}.  Like Bolignano, he
models the four agents $\mathsf{A}$, $\mathsf{B}$, $\Server$ and the spy.
However, his model is more realistic.  $\mathsf{A}$ and $\mathsf{B}$ may
engage in concurrent runs, playing either role; the spy has an identity and a
long-term key.  Lowe has discovered numerous attacks, some of which are
serious~\cite{lowe-fdr,lowe-new,lowe-splice}.  I have found his papers most
useful in developing the Isabelle model.

Meadows's paper on Needham-Schroeder~\cite{meadows-ns} makes direct
comparisons with Lowe's.  She examines the same variants of the protocol and
discusses differences in speed between the NRL Protocol Analyzer and FDR (the
latter is faster).  She reports many other experiments, for example on the
possibility of nonces being compromised.  Not all of these attempts are
successful: sometimes the state space becomes too large.

Her Protocol Analyzer performs an unusual combination of search and proof.
Ostensibly based on brute-force state enumeration, it can also prove by
induction that infinite sets of states are unreachable.  A full analysis
carries the same assurance as a formal proof.  The precise relationship between
Meadows's and my uses of induction needs to be examined.

Schneider~\cite{schneider-verifying}, like Lowe, bases his work
on~CSP~\cite{hoare85}.  But instead of using a model checker, he applies the
laws of CSP in proofs.  A rank function is used to describe how an undesirable
event is prevented.  Proving certain theorems about the rank function
establishes the property in question.  Schneider has published detailed hand
analyses of both the original protocol and Lowe's version.  He considers a
number of authentication properties in increasingly general settings,
ultimately allowing concurrent runs.  Recently, Dutertre and
Schneider~\cite{dutertre-csp-pvs} have mechanized these hand proofs, revealing
many errors in them.  Bryans and Schneider~\cite{bryans-csp} have proved some
simple properties for a single run of the recursive authentication protocol.

Schneider has considered the consequences of allowing messages to satisfy
equational laws on messages.  Many protocols---and attacks!---exploit
algebraic properties of encryption method, particularly RSA~\cite{rsa78}.

\section{Conclusions}\label{sec:conclusions}

The inductive method is simple and general.  We have seen how it handles three
versions of Otway-Rees, two versions of Needham-Schroeder (with public keys),
and a recursive protocol.  The analysis of Needham-Schroeder reveals Lowe's
attack, and I have discovered a new attack in a variant of Otway-Rees.  In
addition to the protocols discussed above, I have analyzed two variants of
Yahalom~\cite{paulson-yahalom}, a simplified version of
Woo-Lam~\cite{abadi-prudent} and the shared-key version of Needham-Schroeder.
Bella and I have looked at Kerberos, which is based on timestamps; its use of
session keys to encrypt other keys complicates its
analysis~\cite{bella-kerberos}.  I have modelled part of the Internet protocol
TLS~\cite{tls-1.0,paulson-tls} in which secret nonces are exchanged, then used
to compute session keys.

Proofs are highly automated.  One Isabelle command can generate thousands of
inferences.  Small changes to protocols involve only small changes to proof
scripts.%
\footnote{Proof scripts are distributed with Isabelle, which can be obtained
  from URL \url{http://www.cl.cam.ac.uk/Research/HVG/Isabelle/dist/}; see
  subdirectory \url{HOL/Auth}.}  %
Analyzing Needham-Schroeder took only 30 hours of my time, the recursive
protocol two weeks.  These figures include time spent extending
the model with public-key encryption and hashing.  Adherence to design
principles such as explicitness~\cite{abadi-prudent} simplifies proofs.

Model checking is an effective means of finding
attacks~\cite{lowe-fdr,lowe-new,lowe-splice}, but it cannot replace theorem
proving.  It copes with only finitely many states, and the failure to find an
attack says nothing about how a protocol works.

An inductive proof is a symbolic examination of the protocol.  Each step is
analyzed in turn.  The reasoning can be explained informally, letting us
understand how the protocol copes with various circumstances.  When a protocol
is modified, the proof scripts for the old version form the starting point for
its analysis.  These scripts take only a few minutes to run, which is
competitive with model checking.

The two methods complement each other.  A protocol designer might use model
checking for a quick inspection and apply the inductive approach to
investigate deeper properties.

Formal methods cannot guarantee security.  
Theorems can easily be misinterpreted.  Needham-Schroeder is correct from
$A$'s point of view, but not from~$B$'s.  The flawed version of Otway-Rees is
correct from the server's point of view, but the other participants cannot
detect tampering (recall \S\ref{sec:further}).  A protocol proof must contain
a separate guarantee---under reasonable assumptions---for each participant.

The attack on the recursive protocol~\cite{ryan-attack} is a sobering reminder
of the limitations of formal methods.  Models idealize the real world: here,
by assuming strong encryption.  Making the model more detailed makes reasoning
harder and, eventually, infeasible.  A compositional approach seems necessary:
different levels of abstraction, such as protocol messages, cryptographic
algorithms, and transport protocols, should be verified separately.
Devising such an approach will be a challenge.

\paragraph*{Acknowledgement}
R. Needham gave much valuable advice.  Thanks are due to J. Bull and D. Otway
for explaining their protocol to me.  A. Gordon suggested a simplification to
the treatment of freshness.  Conversations with M. Abadi, D. Bolignano, G.
Huet, P. Ryan and K. Wagner were helpful.  G. Bella, B. Graham, G. Lowe, F.
Massacci, V. Matyas, M. Staples, M. VanInwegen and several anonymous referees
commented on drafts of this article.  The research was funded by the {\sc
  epsrc} grants GR/K77051 `Authentication Logics' and GR/K57381 `Mechanizing
Temporal Reasoning', and by the \textsc{esprit} working group 21900 `Types'.

\bgroup
\bibliographystyle{plain}\raggedright\frenchspacing
\bibliography{string,atp,funprog,general,isabelle,theory,crossref}
\egroup

\end{document}